\documentclass{aa}

\usepackage{url}
\usepackage{hyperref}
\hypersetup{
    colorlinks = true,
    citecolor = {blue},
}
\usepackage{subcaption}
\usepackage{natbib}
 \bibpunct{(}{)}{;}{a}{}{,} 
\usepackage{multirow}

\setcitestyle{authoryear,open={(},close={)}} 

\usepackage{booktabs}
\usepackage{array}

\usepackage[table]{xcolor}
\definecolor{lightgrey}{rgb}{0.9, 0.9, 0.9}

\usepackage{afterpage}

\usepackage{stfloats} 

\usepackage{placeins}

\usepackage{enumitem,kantlipsum}

\usepackage{xspace}
\newcommand{\fOtwo}{\textit{f}O$_2$\xspace}
\newcommand{\dIW}[1]{$\Delta$IW{#1}\xspace}

\newcommand*\diff{\mathop{}\!\mathrm{d}}

\title{Impact of oxygen fugacity on atmospheric structure and emission spectra of ultra hot rocky exoplanets}

\author{Fabian L. Seidler \inst{1} \and Paolo A. Sossi \inst{1} \and Simon L. Grimm \inst{2}}

\institute{
    ETH Zürich, Department of Earth Sciences, Institute for Geochemistry and Petrology\\
    \email{fseidler@ethz.ch}, \email{psossi@ethz.ch}
    \and
    ETH Zürich, Department of Physics, Institute for Particle Physics and Astrophysics\\
    \email{sigrimm@ethz.ch}
}

\date{April 2024}

\begin{document}

\abstract
{Atmospheres above lava-ocean planets (LOPs) hold clues as to the properties of their interiors, owing to the expectation that the two reservoirs are in chemical equilibrium. Furthermore, such atmospheres are observable with current-generation space- and ground-based telescopes. While efforts have been made to understand how emission spectra can be related to the composition of the lava ocean, the influence of oxygen fugacity has yet to be examined self-consistently.}
{Here we investigate the sensitivity of atmospheric emission spectra of LOPs to key geochemical parameters, namely, temperature ($T$), composition ($\Vec{X}$) and oxygen fugacity (\fOtwo), and the precision to which they can be recovered from observations of hot, rocky exoplanets.}  
{To do so, we consider `mineral' atmospheres produced in equilibrium with silicate liquids. We treat \fOtwo as an independent variable, together with $T$ and $\Vec{X}$ to compute equilibrium partial pressures ($p$) of stable gas species at the liquid-gas interface. Above this boundary, the atmospheric speciation and the pressure-temperature structure are computed self-consistently to yield emission spectra. We explore a wide array of plausible compositions, oxygen fugacities (between 6 log$_{10}$ units below- and above the iron-wüstite buffer, IW) and irradiation temperatures (2000, 2500, 3000 and 3500 K) relevant to LOPs.}
{We find that SiO(g), Fe(g) and Mg(g) are the major species below $\sim$IW, ceding to O$_2$(g) and O(g) in more oxidised atmospheres. The transition between the two regimes demarcates a minimum in total pressure ($P$). Because $p$ scales linearly with $\Vec{X}$, emission spectra are only modest functions of composition. By contrast, \fOtwo can vary over orders of magnitude, thus causing commensurate changes in $p$. Reducing atmospheres show intense SiO emission, creating a temperature inversion in the upper atmosphere. Conversely, oxidised atmospheres have lower $p$SiO and lack thermal inversions, with resulting emission spectra that mimic that of a black body. Consequently, the intensity of SiO emission relative to the background, generated by MgO(g), can be used to quantify the \fOtwo of the atmosphere. Depending on the emission spectroscopy metric of the target, deriving the \fOtwo of known nearby LOPs is possible with a few secondary occultations observed by JWST.}
{}


\maketitle


\section{Introduction}
\label{sec:introduction}

Among the $> 5000$ exoplanets discovered today, there exists a subgroup of $\sim 500$ terrestrial-sized worlds ($R \leq 2 R_\oplus$, $M \leq 10 M_\oplus$) on extremely close-in orbits of the order of a day or less around their star \citep{Zilinskas:2022}. Consequently, they are thought to be tidally locked, with a dayside that experiences permanent illumination and therefore temperatures that may be in excess of the 1 bar solidi of most silicates \citep[$\geq 1500$ K, e.g.,][]{hirschmann2000}. Owing to their extreme characteristics, they are often labeled as Ultra-Short-Period (USP) planets, hot-rocky exoplanets (HREs) or lava-ocean planets (LOPs).

Due to the high temperatures, the silicate surface may not only melt, but also partially vaporise \citep[e.g.][]{Leger:2011, Wolf:2022}, forming a tenuous vapour shroud commonly called a mineral or silicate atmosphere. Chemical thermodynamic equilibrium between the (molten) surface and the atmosphere is expected based on the relative convection and crystallisation timescales of fully molten magma oceans \citep{salvadorsamuel2023} or their surficial counterparts \citep{Kite:2016}, which, for highly irradiated HREs, are long-lived. As a consequence, the atmosphere should carry the imprint of the physicochemical conditions (pressure-temperature-composition) defined at the interface, and therefore - by extension - information on the mantle composition of the planet, making these worlds promising targets for the study of planetary surfaces, interiors and evolution.

To do so, several models have been used to explore hypothetical atmospheric compositions and thermal emission of rocky exoplanets through numerical simulation. \citet{Fegley:1987} introduced the \texttt{MAGMA} code to calculate the partial pressures of stable gas species in high temperature vapours in equilibrium with silicate liquid relevant to the evolution of Mercury. This model was later employed by \citet{Schaefer:2004}, \citet{schaefer2009chemistry}, and \citet{Miguel:2011} for the study of LOPs. According to their findings, alkali metals (Na and K), oxygen (O$_2$ and O), Fe and SiO dominate mineral atmospheres, the specific mixture being a function of irradiation temperature and the composition of the silicate liquid. \citet{Ito:2015} refined this model by extending the aforementioned "zero-dimensional" vapour models to a 1D-atmosphere, incorporating realistic melt thermodynamics based on the MELTS formalism \citep{Ghiorso:1995}, equilibrium atmospheric chemistry, and full radiative transfer. Their work predicts thermal inversions in mineral atmospheres where temperatures steadily increase with altitude, resulting in strong mid-infrared emission features, especially from SiO. \citet{Zilinskas:2022} adopted a similar approach and applied it to known potential LOPs with surface temperatures $>1500$ K, considering a broader range of molecular species (MgO, SiO$_2$, etc.). Their study emphasises the significant influence of TiO and MgO interplay on the thermal emission spectrum for compositions akin to the bulk silicate Earth \citep[BSE, Earth's mantle composition, c.f. ][]{mcdonough1995}, oceanic and continental crust and komatiite, or Mercury's mantle bulk composition; they utilised the original \texttt{MAGMA} code \citep{Fegley:1987, Schaefer:2004}. Their model was later extended to incorporate the major volatile species such as CO$_2$, H$_2$ and H$_2$O \citep{Zilinskas:2022, piette2023rocky}.

However, exoplanet compositions may span a broader range than those examined by \cite{Zilinskas:2022}, which are based on terrestrial rocks. Particularly, some rocky exoplanets might have iron (FeO)-enriched mantles by H$_2$O-induced oxidation of metal \citep{ElkinsTanton_2008} or high oxygen abundances in the planetary source material \citep{Wang:2019}. The size of the metallic core is also degenerate with the FeO-content of the mantle when only mass and radius are considered \citep{Dorn:2015}, highlighting one key uncertainty: the fraction of oxygen condensing during planet formation \citep{Putirka:2019}. Small amounts of available oxygen might lead to planets rich in metallic iron and thus large cores, whereas overabundant oxygen could convert more metallic iron into FeO, which would reside in the mantle \citep{wang_2018_enhanced_constraints, wang_2022_detailed_chemical_compositions_II, Spaargaren_2023}. This exchange is governed by the chemical availability (chemical potential) of oxygen (assuming the Gibbs free energy of formation of pure O$_2$ is 0 at 1 bar and $T$; its standard state):
\begin{equation}
    \label{eq:fO2}
    \mu_{O_2} = RT \ln{ fO_2}
\end{equation}
where $R$ is the gas constant, $T$ the temperature and \fOtwo the oxygen fugacity, a measure for the effective partial pressure of O$_2$, $p\text{O}_2 = \varphi_{O_2} \cdot$ \fOtwo, where $\varphi$ is the fugacity coefficient. For an ideal gas, $\varphi=1$, such that fugacity and partial pressure are equivalent; at higher total pressure ($P>1000$ bar), they start to diverge markedly. Taking the above results together, one of the key uncertainties in constraining terrestrial exoplanet composition are \textit{i)} the bulk elemental abundances of the major rock forming elements (Fe, Mg, Si) and \textit{ii)} the \fOtwo, as approximated by the FeO/Fe ratio of the planet. The \fOtwo at the surface is crucial in shaping the nature of potential atmospheres owing to both the stoichiometry of vaporisation reactions \citep{visscherfegley2013,sossi2019,Jaeggi:2021,Wolf:2022} and homogeneous gas phase reactions \citep{sossifegley2018,Sossi:2020:redox_atmo}. 

The aforementioned studies either assume a fixed, arbitrary \fOtwo \citep{piette2023rocky} or rely on codes that try to predict the partial pressure of O$_2$ in the vapour from the stoichiometry of the vaporisation reactions themselves; these codes include the original \texttt{MAGMA} code \citep{Fegley:1987,Schaefer:2004, visscherfegley2013} and the recently published \texttt{LavAtmos} code \citep{vanBuchem:2023}.

However, the \fOtwo calculated according to the stoichiometric method is correct only when \textit{a)} all possible reactions that release O$_2$ in the vapour phase are considered and \textit{b)} the set of melt components, and their thermodynamic properties, is complete and accurate (due to the dependence of \fOtwo on melt activities, see Sec. \ref{methods:vapour}). The \texttt{MAGMA} and \texttt{LavAtmos} codes are concerned only with Na, K, Ti, Fe, Ca, Al, Mg, Si self-consistently, yet any other element, for example, chromium, sulfur, carbon or hydrogen, may also be present in natural silicate liquids. In this case, the \fOtwo given by the stoichiometric approach would be incorrect. Under thermodynamic equilibrium, the chemical potential, $\mu_{O_2}$, must be the same in the atmosphere as in the melt, and hence, there is a singular \fOtwo that defines the system (Eq. \ref{eq:fO2}). The \fOtwo of a magma ocean-atmosphere system is a function of its bulk composition at a given pressure and temperature \citep[e.g.,][]{Sossi:2020:redox_atmo,Hirschmann:2022}. However, the compositions of exoplanets are not known, such that the \fOtwo cannot be predicted a priori.  Therefore, here, we consider \fOtwo as an independent variable, which resolves both problems.

We aim to study the impact of (hypothetical) exoplanet mantle compositions as well as of oxygen fugacity on the formation, nature and detectability of atmospheres on lava ocean planets. Our new model is developed in a vein similar to that described by \citet{Ito:2015}, \citet{Zilinskas:2022} and \citet{piette2023rocky}, including self-consistent thermodynamics and its effect on atmospheric structure and spectra. This study is structured as follows: Section \ref{sec:methods} details the methodology; this entails the description of the modified \texttt{MAGMA} code that permits the oxygen fugacity to be controlled as an independent variable, the derivation of the atmospheric speciation, and the atmospheric pressure-temperature structure and emission spectrum. This model is applied to various hypothetical compositions, which we also derive in this section (Sec. \ref{methods:exoplanet_compositions}). Section \ref{sec:results} shows the results. Section \ref{sec:discussion} discusses the implications, and we conclude and summarise in Section \ref{sec:summary}.


\section{Methods}
\label{sec:methods}

\subsection{Vaporisation of the lava ocean}
\label{methods:vapour}

The first step in modelling a mineral atmosphere is the derivation of the equilibrium pressure and vapour composition at the ocean-atmosphere interface. The initial temperature of the magma ocean is estimated to be that of the dayside temperature of a tidally locked black-body planet, but it is adjusted in subsequent iterations to match the atmospheric base temperature (see Sec. \ref{methods:radiative_transfer} and \ref{methods:temperature}). To then simulate its vaporisation, we use a modified version of the MAGMA code \citep{Fegley:1987, Schaefer:2004} that allows us to fix the oxygen fugacity (\fOtwo) as a free parameter, rather than it being determining via stoichiometric vaporisation, as done hitherto. We use the equilibrium constants from \citet{Fegley:1987}, including for the reaction $SiO_2(l) = Si(g) + 2O$, despite it having been revised later \citep{Schaefer:2004}, because the \cite{Schaefer:2004} model underestimates partial pressures of Si-bearing gases in an attempt to improve those for Na and K (which we omit unless otherwise mentioned, due to reasons detailed in Sec. \ref{methods:exoplanet_compositions}). We further add the reactions
\begin{subequations}
\label{eq:fO2_buffers}
    \begin{align}
        &\text{FeO}(l, silicate) = \text{Fe}(l, metal) + \frac{1}{2}\text{O}_2 \label{eq:IW_buffer}\\
        &2 \text{FeO}(l, silicate) + \frac{1}{2}\text{O}_2 = \text{Fe}_2\text{O}_3(l, silicate) \label{eq:WH_buffer}
    \end{align}
\end{subequations}

that describe the conversion of the FeO component dissolved in the silicate liquid (l, silicate) to account for the change in silicate melt composition due to effects of \fOtwo. This leads to \textit{a)} the saturation of metallic Fe at low \fOtwo as a separate liquid phase (l, metal) and \textit{b)} the presence of Fe$^{3+}$ in silicate liquids promoted under oxidising conditions \citep[e.g.,][]{KressCarmichael1991, BerryONeill2021}. The restriction of fixed \fOtwo marks the chemical network as an open system with respect to O$_2$, while the number of atoms of non-oxygen elements (Si, Mg, Fe, and so on) is conserved\footnote{The total oxygen content of the melt is dependent on \fOtwo. Since we allow only iron in multiple oxidation states (Fe, FeO, Fe$_2$O$_3$), their relative abundances will change in response. In realistic systems however, the situation is reversed: the composition sets the \fOtwo. The use of fixed oxygen fugacity therefore requires the assumption of a "nearly" infinite melt reservoir (compared to the vapour mass), otherwise the open system assumption is violated.}.
The respective equilibrium constants for reactions \ref{eq:fO2_buffers} were derived from the available reactions in \texttt{MAGMA} so as to remain self-consistent.

In \texttt{MAGMA}, a series of reactions between assumed melt components ("pseudocompounds") is balanced to iteratively solve for the activity coefficients of the parent oxide species. The activity of a species in the melt is defined as
\begin{equation}
    \label{eq:activity}
    a_i = x_i \gamma_i
\end{equation}
and relates the mole fraction of component $i$, $x_i$ in the melt to its activity coefficient $\gamma_i$, which is any real positive number. Unfortunately, the implementation limits the range of all $\gamma_i$ between 0 and 1, whereas, in reality, some activity coefficients (particularly for FeO) might exceed unity \citep{Wolf:2022}. We further fix the activity of Fe(l, metal), when stable, at unity, $a_{Fe}=1$ because it constitutes its own phase due to its insolubility in silicate liquids. The \fOtwo below which metallic iron can form is computed by exploiting Eq. \ref{eq:IW_buffer}, re-arranged to solve for the oxygen fugacity: 
\begin{equation}
    \label{eq:transition_fo2}
    f\text{O}_2=\frac{K \cdot a_{\text{FeO(l, silicate)}}}{a_{\text{Fe(l, metal)}}}
\end{equation}
where $K$ is the equilibrium constant of reaction \ref{eq:IW_buffer}. \texttt{MAGMA} computes $\gamma_{FeO}$ in the silicate; since $a_{FeO}$ is set by Eq. \ref{eq:transition_fo2}, we can invert Eq. \ref{eq:activity} for $x_{FeO}$ (in the silicate) to solve for the amount (i.e., mole fraction) of metallic Fe (in the metal) with the additional mass balance constraint;
\begin{equation}
    \label{eq:mass_balance}
    \Sigma \mathrm{Fe} = \mathrm{Fe}^0 + \mathrm{FeO} + \mathrm{Fe_2O_3}
\end{equation}
where the ratio $x$Fe$_2$O$_3$/$x$FeO is also fixed at a given $P$, $T$ and \fOtwo by eq. \ref{eq:WH_buffer}. Note that this is incorporated into the iterative scheme of \texttt{MAGMA} in order to account for the decrease of $x$FeO$(l, silicate)$, and, correspondingly, an increase in the mole fractions ($x$) of all other oxides. A detailed example of how \fOtwo affects the thermochemistry of the system is shown in Appendix \ref{apx:bse_fugacity_series:thermochemistry}. However, some caveats still remain: firstly, in natural systems, other elements can be incorporated into the metal phase, e.g. Si or O, which are not considered here; we discuss the implications in Sec. \ref{discussion:degeneracy_between_meltcomp_and_fO2}. Secondly, we implicitly assume that the metal phase remains well-mixed with the silicate phase, and hence $p$Fe is buffered at a given pressure and temperature according to the reaction Fe(l, metal) = Fe(g), where $a$Fe(l, metal) = 1. Were the metal to separate, then the silicate melt would record the \fOtwo set at the last pressure-temperature of equilibration with the metal. Any subsequent changes in \fOtwo would be dictated by Eq. \ref{eq:WH_buffer}.

At constant temperature, the formation of a gas species $M_\mu O_\nu(g)$ (where M refers to any metal species, and $\mu$ and $\nu$ are the stoichiometric coefficients) involves two types of reactions. The first is the heterogeneous vaporization of an oxide component $M_m O_n(l)$ from the silicate liquid, occurring between two phases - melt and gas \citep[cf.][]{sossi2019}:

\begin{equation}
\label{eq:heterogeneous_equation}
M_m O_n(l) = m M(g) + \frac{n}{2} O_2(g)
\end{equation}

The second type is a homogeneous reaction, i.e. occurring within a single phase (gas):

\begin{equation}
\label{eq:homogeneous_equation}
M_\mu O_\nu(g) = \mu M(g) + \frac{\nu}{2} O_2(g)
\end{equation}

Taken together, the fugacity $fM_\mu O_\nu$ of a gas species $M_\mu O_\nu$ is:

\begin{equation}
    \label{eq:f_of_gas_species}
    fM_\mu O_\nu = \frac{K_{het}^\alpha}{K_{hom}} \cdot a^\alpha_{M_mO_n(l)} \cdot f\text{O}_2^{\beta}
\end{equation}

where we define $\alpha :=\frac{\mu}{m}$ and $\beta :=\frac{1}{2}(\nu - \mu \frac{n}{m})$; $K_{\text{het}}$ and $ K_{\text{hom}}$ are the respective equilibrium constants to Equations \ref{eq:heterogeneous_equation} and \ref{eq:homogeneous_equation}, adapted from \texttt{MAGMA}, and 
$a_{M_mO_n(l)}$ is the aforementioned activity. \fOtwo is kept as a free parameter. 
The coefficients for evaporation of all gas species considered in this study are given in Table \ref{tab:coeffs}.

\begin{table}[!t]
\caption{Coefficients of gas species formation (Eq. \ref{eq:f_of_gas_species}).}
\centering
\begin{tabular}{lll}
\toprule
  Gas species                                                                         & $\alpha$   & $\beta$    \\
\midrule
\rowcolor{lightgrey}
\begin{tabular}[c]{@{}l@{}}SiO$_2$, MgO, FeO, TiO$_2$, CaO, Na$_2$O, K$_2$O\end{tabular} & 1   & 0    \\
SiO, TiO, Mg, Fe, Ca, Al$_2$O$_2$                                                           & 1   & -1/2 \\
\rowcolor{lightgrey}
Si, Al$_2$O                                                                              & 1   & -1   \\
Al                                                                                    & 1/2 & -3/4 \\
\rowcolor{lightgrey}
Na, K, AlO                                                                          & 1/2 & -1/4 \\
NaO, KO                                                                               & 1/2 & 1/4  \\
\rowcolor{lightgrey}
O$_2$                                                                                 & -   & 1 \\
O                                                                                     & -   & 1/2 \\
\bottomrule
\end{tabular}
\label{tab:coeffs}
\end{table}

Finally, we impose the ideal gas law in order to completely define the fugacity, $f\text{G}$, of any gas $\text{G}$, that is $f\text{G}=p$G, with $p$ being the partial pressure. This is justified by the anticipated small value of the total vapour pressures \citep[$\leq$ 1000 bar,][]{Ito:2015, Zilinskas:2022, Wolf:2022, vanBuchem:2023}, for which fugacity and pressure coincide.

The \fOtwo is itself a function of temperature and pressure, and it is hence useful to normalise absolute \fOtwo to a mineral buffer; here, we choose the iron-wüstite (IW) buffer from \citet{Hirschmann:2008} since the main redox control occurs via FeO:
\begin{equation}
    \label{eq:IW-buffer}
        \log f\text{O}_2 = \text{IW}(P, T) + \Delta \text{IW}
\end{equation}
We denote the (logarithmic) deviation from IW (which is Eq. \ref{eq:IW_buffer} but for a reaction of pure solid Fe and FeO) as $\Delta$IW. Technically, Eq. \ref{eq:IW_buffer} is pressure dependent, but the anticipated atmospheric pressures of $\leq$ 1000 bar have a negligible influence on the volume change - and therefore the pressure-sensitivity - of solid-solid buffers.


\subsection{Atmospheric speciation}

\label{methods:atmospheric_speciation}

Far above the lava ocean, chemical and thermal equilibrium with the underlying magma ocean is lost, and, as such, the temperature and pressure in the atmosphere differ from those at the interface. Here, we assume that all atmospheric layers have the same bulk elemental composition as the vapour in equilibrium with the melt, i.e., the atmosphere is assumed to be well-mixed, even though this might not be the case in the presence of expected thermal inversions \citep[cf.][]{Ito:2015}. Further, the condition of thermochemical equilibrium in each atmospheric layer is invoked.
Following these restrictions, we can compute the gas speciation with the \texttt{FastChem} code \citep{Stock:2018, Stock:2022}, which produces look-up tables for the thermodynamic properties of gas species, spanning a range of $10^{-8}-10^3$ bar and 1500-6000 K (sufficient to model mineral atmospheres). The reactions involved in \texttt{FastChem} are displayed in Table \ref{tab:fastchem_reactions}. Although reaction kinetics are ignored, the high temperatures ($\geq$ 2000 K) are likely sufficient to establish equilibrium. For this study, we neglect the effect of photochemistry.

\subsection{Opacities}
\label{methods:opacities}

\begin{figure*}[!t]
    \centering
    \includegraphics[width=\linewidth]{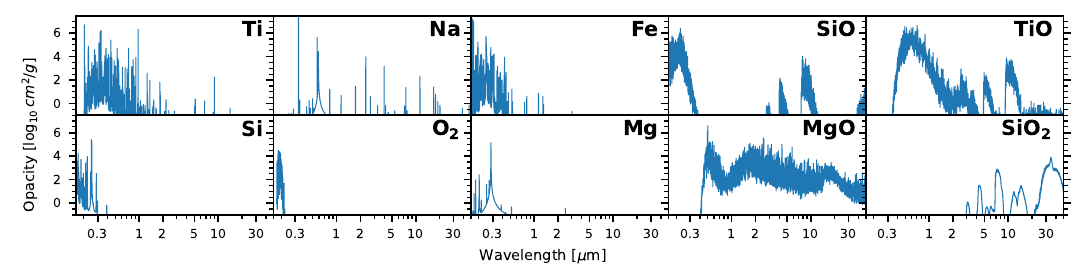}
    \caption{Opacities of major gas species sampled at $\Delta \nu$=0.01 cm$^{-1}$, $\nu=\frac{c}{\lambda}$, at 2500 K and 0.1 bar (a typical value for the photosphere) as a function of wavelength. Note that the spectral lines of atomic O are omitted since are so weak that they would not be visible in the plot.}
    \label{fig:molecular_opacities}
\end{figure*}

\subsubsection{Atoms}
\label{methods:opacities:atoms}

The atomic species allowed in the modelled atmospheres are Si, Mg, Fe, O and Ti (and, in a few cases, Na \& K). Their opacities are obtained from The Data \& Analysis Center for Exoplanets (DACE)\footnote{\url{https://dace.unige.ch/opacityDatabase/}, please refer to the acknowledgements or \url{https://dace.unige.ch/daceTeam/} for more details.}; from there, we obtain the tabulated opacities as a function of wavelength. We utilise the line lists from \citet{Kurucz:2017} for all atomic species. The spectra on DACE were generated with the opacity calculator \texttt{HELIOS-K} \citep{Grimm:2015, Grimm:2021}, using a resolution of $\Delta \nu = 0.01~\text{cm}^{-1}$. The lines are treated as Voigt-profiles where the line wings are cut at a distance of 1000000 $\Delta \nu~\text{cm}^{-1}$ from the line center, effectively amounting to no line cut. This procedure almost certainly overestimates the effect of the line wing on the opacity, but the strong resonant lines of many metal gases do not merit the assumption of line cuts (e.g., Mg at 285 nm). Ideally, they would need to be modelled with other line shape theories, such as done for Na in \cite{Zilinskas:2022}. We use atomic opacities only for a pressure of $10^{-8}$ bar, noting that their opacities on DACE are only available at $10^{-8}$ bar. 

\subsubsection{Molecules}
\label{methods:opacities:molecules}

We include the molecules SiO \citep{Yurchenko_2021_SiOUVenIR}, SiO$_2$ \citep{Owens_2020_SiO2_OTY3}, MgO \citep{Li_2019_MgO_LiTY}, O$_2$ \citep{Gordon_2017_Hitran}, TiO \citep{McKemmish_2019_TiO_Toto}. Their opacities were obtained from the Exoclimes simulation platform\footnote{\url{https://chaldene.unibe.ch/}; by the time of publication of this paper, its contents will be available from DACE, from where it should be preferentially obtained.} which hosts opacities prior to their upload to DACE. The line lists for most molecules stem from ExoMol\footnote{\url{https://www.exomol.com/}} \citep{Tennyson:2016}, except for O$_2$, which originates from HITRAN\footnote{\url{https://hitran.org/}} \citep{Rothman:2005, Rothman:2009, Gordon_2017_Hitran, Gordon:2022}. Contrary to the atoms, the Voigt profiles have a line wing cut at 100 cm$^{-1}$ from the line core (smaller than for the atomic species, as they lack broad resonance lines), but are similarly sampled at a resolution of 0.01 cm$^{-1}$. Most molecular species here have line list data available over the range 1500-4500 K, except for SiO$_2$, for which no data beyond 3000 K exist.

Further, we use opacities sampled over a pressure range 10$^{-8}$ -- 10$^3$ bar. Pressure induces an increase in the width of a spectral line due to collisions with the surrounding atmosphere, an effect known as pressure broadening which becomes more relevant for atmospheres of higher pressure. When approximating line shapes with Voigt profiles, the pressure effect enters through an increase in the half-width $\Gamma_L$ of the line \citep[c.f.][]{Grimm:2021}. ExoMol models the half-width as \citep{Tennyson:2016}:
\begin{equation}
    \Gamma_L \propto \Gamma_{ref} \left( \frac{T_{ref}}{T} \right)^n \left( \frac{P}{P_{ref}} \right)
\end{equation}
where $\Gamma_{ref}$ and $n$ are the broadening parameters and depend on the nature of the ambient gas. Different lines of the same species might have different coefficients. The broadening coefficients that are available are usually valid only for interactions with H$_2$, He or air as background gases. However, for the metal- and metal oxide species relevant to this study, no broadening coefficients - regardless of atmospheric background - are available. We therefore make do with the existing default parameters from ExoMol, $\Gamma_{ref}=0.07$ cm$^{-1}$/bar and $n=0.5$, which are applied to all lines of all molecules indiscriminately.

O$_2$ is treated differently as it originates from the HITRAN database. In this case, the line broadening is assumed to stem from collisions between molecules of the same species, and the half-width takes the following form:
\begin{equation}
    \Gamma_L \propto \Gamma_{ref} \left( \frac{T_{ref}}{T} \right)^n \left( \frac{ \Gamma_{self} P_{self}}{P_{ref}} \right)
\end{equation}
with the broadening parameter $\Gamma_{self}$ for self-broadening, the reference pressures and temperatures $P_{ref}$ and $T_{ref}$ as well as the temperature exponent $n$.

An example for the opacities of major gas species in mineral atmospheres, sampled at 2500 K and 0.1 bar (broadening included) is shown in Fig. \ref{fig:molecular_opacities}.

We acknowledge that our treatment of line broadening is a simplification necessitated by a lack of data. However, we also consider a case with and without broadening and highlight the differences in the computed spectra (cf. Sec. \ref{sec:assumptions_and_limitations} and Appendix \ref{app:broadening}).

\subsection{Radiative transfer}
\label{methods:radiative_transfer}

The atmospheric structure is computed self-consistently with the radiative transfer code \texttt{HELIOS} \citep{Malik:2017}\footnote{\url{https://github.com/exoclime/HELIOS}}. This code treats the atmosphere on a log$_{10}$-equidistant grid confined between the pressure at the top-of-the-atmosphere (TOA), fixed at $10^{-8}$ bar, coinciding with the lower limit on the available opacities, see Sec. \ref{methods:opacities}, and the bottom-of-the-atmosphere (BOA), which equals the vapour pressure atop the lava ocean. 

\texttt{HELIOS} uses the tabulated atmospheric speciation and the specified opacities to construct each layer's transmissivity $\mathcal{T}$. The mixing of the opacity of each species within the 0.01 cm$^{-1}$ wavelength-wide spectral bin is calculated via the random overlap method \citep{lacis1991description} unless the maximum opacity of the species to be added is 1\% of the total opacity, in which case the correlated-k method is used \citep{Malik:2017}. Effectively, this assumes that the transmissivity is a product of the individually opacities weighted by the mixing ratio of the respective gas species \citep{Amundsen:2017}. $\mathcal{T}$ is used to solve the radiative transfer problem by iteratively adjusting the temperature profile to the radiation flux \citep[cf. ][]{Malik:2017}, starting from an isothermal atmosphere as the initial guess.

Further boundary conditions are the stellar spectrum at TOA (which can either be a black-body or a (user-provided) real/simulated spectrum, either of which is scaled to match an arbitrary imposed irradiation temperature $T_{\text{irr}}$, which we cover in detail in Sec. \ref{methods:temperature}), and the internal temperature which dictates the heat flux, $F$ at the BOA. The internal temperature, $T_{int}$, is given by:

\begin{equation}
T_{int} = \left(\frac{F_{out}-F_{in}}{\sigma}\right)^{\frac{1}{4}}
    \label{eq:T_int}
\end{equation}

where $\sigma$ is the Stefan-Boltzmann constant, and reflects the energy balance between that released by the planet to space ($F_{out}$) and that being shed from the interior ($F_{in}$). We leave this parameter at its default setting of 30 K. We neglect any convective adjustment to the temperature-pressure profile, since previous studies have found the atmosphere to be characterised by a thermal inversion and is thus strongly stratified \citep{Ito:2015, Zilinskas:2022}.

The converged solution yields the pressure-temperature profile of the atmosphere. Various useful byproducts are generated, such as the spectrum and atmospheric transmission function. However, the BOA-temperature changes during the run, hence the melt temperature would no longer correspond to the initial guess, i.e. the substellar temperature, and is iterated accordingly. This requires subsequent calls to both the vaporisation routine (Section \ref{methods:vapour}), \texttt{FastChem} and \texttt{HELIOS}, until the temperature of the melt has converged, at which point the equilibrium partial pressures of the gases species stabilise and the atmosphere becomes stationary. In our case, we consider $\Delta T \leq 35$ K acceptable, given the wide range of temperatures considered. Upon convergence, the surface pressure may have changed by an order of magnitude from its initial guess. However, the irradiation temperature of the planet is maintained throughout a run.

\subsection{Temperature}
\label{methods:temperature}

The energy balance of the planet is set by the stellar radiation input, which predominately depends on the orbital separation $d$. \texttt{HELIOS} uses the stellar spectrum and $d$ to infer the radiation flux at TOA. However, describing the problem via the planet's irradiation temperature $T_{\text{irr}}$ instead is more intuitive, and generalizes the result for planets that orbit larger or smaller stars of sufficiently similar spectrum. We find the orbital separation $d$ which matches the desired $T_{\text{irr}}$ by:
\begin{equation}
    \label{eq:orb_sep}
    d = \sqrt{\mathfrak{f} \cdot (1-A_B)} R_\star \left(\frac{T_\star}{T_{\text{irr}}} \right)^2
\end{equation}
where $A_B$ is the bond albedo \citep[assumed to be 0, i.e., the molten surface is a perfect black body,][]{essack2020albedo} and $T_\star$ and $R_\star$ the effective temperature and radius of the host star (note that here we also assume the star to be a black body, which is an acceptable assumption to derive $d$). $\mathfrak{f}$ is the dilution factor, a technical necessity to account for any redistribution of energy due to geometrical or dynamical processes. A tidally locked planet with no heat redistribution requires $\mathfrak{f}=\frac{2}{3}$, imposed by the spherical geometry of the planet; such an approximation is warranted for LOPs based on their expected thin atmospheres with little heat redistribution \citep{Nguyen:2021}, however, it relies on the parallel beam approximation for the incident stellar light, which cannot be wholly descriptive for planets as close-in as LOPs \citep[eg.][]{Leger:2011, carter2018estimation, Nguyen:2021}. Luckily, the choice of $\mathfrak{f}$ is not of relevance for our study of 1D-atmospheric profiles parameterised by $T_{\text{irr}}$, unless otherwise mentioned. We refer the interested reader to \citet{Hansen:2008} for a more in-depth discussion of $\mathfrak{f}$.


\begin{figure}[!t]
    \centering
    \includegraphics[width=\linewidth]{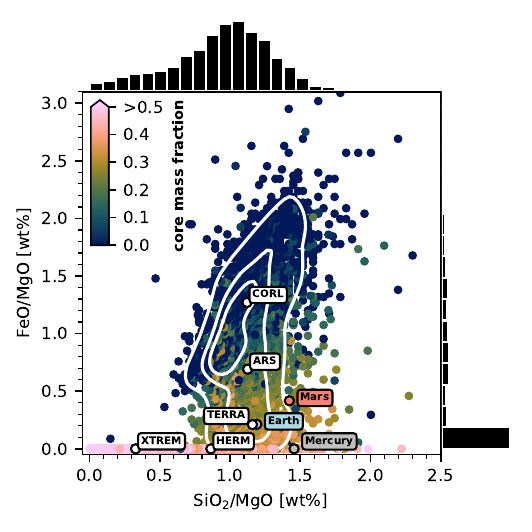}
    \caption{Spread in SiO$_2$/MgO and FeO/MgO ratios in the mantles of prospective exoplanet compositions according to the Hypatia database. See section \ref{methods:exoplanet_compositions} for modelling details. Contours denote the distribution withing the planets that contain FeO in their mantles, and define the levels where the distribution contains 16\%, 50\% and 84\% of probability mass (from inner to outer contour). The white dots represent hypothetical exoplanet compositions selected by a Gaussian Mixture Model. The mantle ratios SiO$_2$/MgO and FeO/MgO are shown for the solar system planets Earth \citep{mcdonough1995}, Mars \citep{khan2022geophysical} and Mercury \citep{nittler2017chemical} (Venus is assumed to be similar to Earth).
    }
    \label{fig:archetypes}
\end{figure}

\begin{table}
\centering
\caption{Hypothetical terrestrial exoplanet compositions.}
\label{tab:archetypes}
\begin{tabular}{rrrrrr}
\toprule
{} & CORL & ARS & TERRA & HERM & XTREM \\
\multicolumn{1}{l}{\textit{mantle}} & & & & & \\
\midrule
SiO$_2$ & 31.26 & 37.15 & 44.89 & 41.54 & 21.30 \\
FeO & 35.43 & 22.82 & 7.60 & 0.00 & 0.00 \\
Al$_2$O$_3$ & 2.68 & 3.37 & 4.01 & 5.39 & 7.81 \\
CaO & 2.62 & 3.51 & 4.06 & 4.72 & 5.56 \\
MgO & 27.89 & 33.01 & 38.72 & 48.11 & 64.97 \\
TiO$_2$ & 0.12 & 0.15 & 0.18 & 0.24 & 0.35 \\
\multicolumn{1}{l}{\textit{core}} & & & & & \\
\midrule
Fe &  & 100 & 100 & 93 & 74 \\
Si &  & 0 & 0 & 7 & 26 \\
\midrule 
CMF & 0 & 18 & 29 & 39 & 46 \\
\bottomrule
\multicolumn{6}{p{0.95\linewidth}}{Notes. The oxides SiO$_2$, MgO, FeO, Al$_2$O$_3$, CaO and TiO$_2$ comprise the mantle. The core contains Fe and Si and has a mass fraction of $\text{CMF} := \frac{M_\text{core}}{M_\text{core} + M_\text{mantle}}$ ("core mass fraction") with respect to the planet. All quantities are in wt\%.}
\end{tabular}
\end{table}

\subsection{Exoplanet compositions}
\label{methods:exoplanet_compositions}

The compositions of hypothetical exoplanetary mantles were estimated from stellar compositions from the Hypatia catalogue \citep{Hinkel2018}. To do so, their elemental abundances (in dex) were converted to elemental fractions. The bulk abundances of the major elements Si, Fe, Al, Ca, Mg and O were then scaled to the observed Sun-to-Earth ratios \citep[e.g.,][]{Wang:2019} to mimic fractionation attending planetary accretion. Oxygen is then stoichiometrically distributed among the elements in the order Ca-Al-Mg-Si-Fe, based on the order of their respective metal-metal oxide equilibrium constants at 1 bar. The amount of O might be insufficient to oxidise all Fe and Si, in which case they will form the metallic core. However, this model is merely an approximation since the exact partitioning of O, Si and Fe between mantle and core depends on the pressure and temperature of equilibration \citep{rubie2011heterogeneous}, which, for exoplanets, is unknown.

This procedure results in 5296 planets, too many to model individually. Therefore, we use a Gaussian Mixture Model (GMM) to extract representative exoplanetary mantle and core compositions (see Appendix \ref{apx:GMM}). To the oxides listed above we add TiO$_2$ due to its strong impact on the atmospheric structure and spectrum \citep{Zilinskas:2022}. Given that Ti is a refractory element with a similar condensation temperature to Al \citep{lodders1998planetary}, we adopt a constant TiO$_2$/Al$_2$O$_3$ ratio as is observed in BSE: $\sim 0.0578$ by molar abundance or $\sim 0.0451$ by mass \citep{mcdonough1995}. The constructed planetary compositions are listed in Table \ref{tab:archetypes}, and their position in the exoplanetary compositional space is indicated in Figure \ref{fig:archetypes}.

We neglect Na$_2$O and K$_2$O in this analysis, as they are not refractory species, and hence there is no orthodox way to estimate their abundance in exoplanet compositions; they are treated separately (see Sec. \ref{discussion:alkali}).


\section{Results}
\label{sec:results}

We explore the possible parameter space of hot, rocky exoplanet (HRE) atmospheres on a grid with 3-axes representing composition, temperature, and oxygen fugacity. For the composition, we use the planetary endmembers listed in Table \ref{tab:archetypes}, and assume they represent fully molten magma oceans. \fOtwo covered the range $\Delta \text{IW} \in \left(-6, -4, -2, 0, 2, 4, 6\right)$; some 12 orders of magnitude. This range is motivated by the trend observed among Solar System objects; the present-day upper mantle of the Earth has \dIW{+3.5} \citep{Frost:2008_redox_mantle}, Mars is roughly at \dIW{}, whereas Earth and Mercury are inferred to have undergone core formation at \dIW{-2} and \dIW{-5} respectively \citep{righter2006}.
The irradiation temperatures were set to 2000, 2500, 3000 and 3500 K, which covers the temperature range of HREs \citep[cf.][]{Zilinskas:2022}. In total, this amounts to a grid of $5\times7\times4=140$ HRE atmospheres. We further assume a $1 M_\oplus$ and $1 R_\oplus$ planet, orbiting a $1 R_\odot$ star. The stellar spectrum is that of the Sun \citep{Gueymard:2003}, but beyond 2.9 {\textmu}m, the spectrum is assumed to be a black body of 5770 K temperature in order to remove erratic noise. The extrapolation was performed with the \texttt{startool} program, which is part of the \texttt{HELIOS} package \citep{Malik:2017}. Most stars with potential lava ocean planets are "Sun-like" FGK stars \citep{Zilinskas:2022}.

\subsection{Vapour chemistry}
\label{results:vapour_chemistry}

\begin{figure}[!t]
    \centering
    \includegraphics[width=\linewidth]{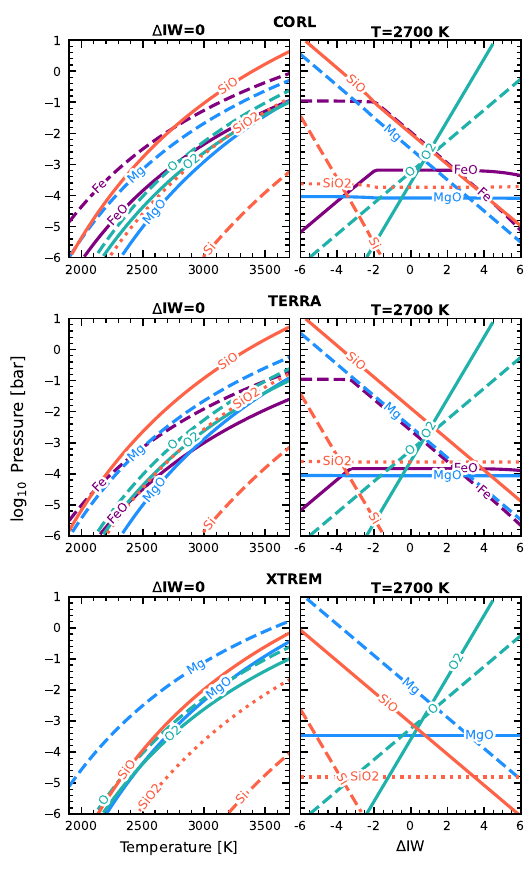}
    \caption{Vapour composition directly above the magma ocean as a function of temperature (left) and oxygen fugacity (right) for three representative mantle compositions in our sample: the composition for a coreless planet, CORL (top), the Earth-analogue TERRA (center) and the extremely silicon-depleted XTREM (bottom); see Table \ref{tab:archetypes} for the respecitve compositions.}
    \label{fig:melt_vapor_evolution}
\end{figure}

\subsubsection{Effect of temperature}
\label{results:vapour_chemistry:temp}

Temperature dictates the total amount of outgassing; the pressure of any gas species increases exponentially with temperature. Because all evaporation reactions have similar enthalpies (curves are approximately parallel in Fig. \ref{fig:melt_vapor_evolution}, left column, see also \cite{sossifegley2018}, relative changes in the stabilities of gas species are minor, but not negligible. As shown by \citet{schaefer2009chemistry, Miguel:2011, Schaefer_2012_vaporization_of_Earth}, we find that low-temperature atmospheres ($\sim$ 2000 K) are rich in Fe and some Mg and SiO. FeO may dominate over MgO (and under oxidising conditions over SiO as well) when the melt is comparatively cold and sufficiently iron enriched (a few wt\%). At higher temperatures ($\geq 2500$ K), atmospheres become more SiO-rich at the expense of Fe, irrespective of composition (provided iron is present, archetypes CORL, ARS and TERRA), even when the FeO content of the mantle is strongly enhanced ($\sim 35$ wt\%, as in CORL).

\subsubsection{Effect of composition}
\label{results:vapour_chemistry:comp}

Unlike temperature, the effect of composition on the outgassed atmospheres is, to a first-order, linear, where $p_i \propto a_i$ (cf. eq. \ref{eq:f_of_gas_species}). Hence, as the mole fractions of the major oxide components in exoplanetary mantle compositions typically vary by a factor 2 -- 3 (Table \ref{tab:archetypes}), the partial pressures of the gas species containing a given element vary by the same factor. This neglects any potential change in the activity coefficient ($\gamma$; eq. \ref{eq:activity}) with varying bulk melt composition, but this is a secondary effect in all cases \citep[see][for more discussion]{Wolf:2022}.
Following this logic, since most of the planetary archetypes have similar Mg/Si ratios, the equilibrium partial pressures - and hence the total abundances - of their respective gas species (Si, SiO, SiO$_2$, Mg, MgO) are nearly identical. The only exceptions are compositions that are extremely deficient in SiO$_2$, such as XTREM. In this case, $p$SiO is second only to $p$Mg in the vapour for the temperature range considered ($\leq 3500$ K).
The abundance of iron, as FeO, on the other hand differs among our compositions, and thus the partial pressures of Fe-bearing species vary between high-FeO (CORL) and low-FeO end-members (TERRA) by a factor of roughly 2-3, as expected. FeO-absent planetary endmembers (HERM \& XTREM) have no Fe or FeO in their atmosphere. 

\subsubsection{Effect of oxygen fugacity}
\label{results:vapour_chemistry:fO2}

Unlike the effect of temperature or composition, \fOtwo can significantly change the nature and abundance of the outgassed atmosphere, by virtue of the fact that it varies by orders of magnitude. The most prominent effects are listed below. 

\begin{itemize}
    \item The partial pressures of SiO, Mg and Fe scale as $\log p_i \propto -\frac{1}{2} \log f\text{O}_2$. Hence, their partial pressures are low for highly oxidised atmospheres ($\Delta \text{IW}>3.5$), but predominate under reducing- to highly reducing conditions ($\Delta \text{IW} < 0$). Their relative proportions, however, are nearly independent of \fOtwo, because their vaporisation reactions are defined by the same stoichiometry (cf. Eq. \ref{eq:f_of_gas_species}, Table \ref{tab:coeffs}).
    
    \item The partial pressures of MgO(g), SiO$_2$(g) and FeO(g) - the latter with limitations that are discussed below - are independent of \fOtwo. Hence, they are less abundant than their oxygen-poor counterparts; Mg(g), Si(g), SiO(g), Fe(g) for highly reducing to oxidising conditions ($\Delta \text{IW}<3.5$, Fig. \ref{fig:melt_vapor_evolution}) but become the dominant form of the respective element in the gas phase at highly oxidising conditions ($\Delta \text{IW} > 3.5$). Their partial pressures relative to other gas species (i.e., O$_2$ and O), however, remain relatively low at these conditions.
    
    \item The partial pressure of FeO(g) declines as FeO(l) is reduced to its metallic form in the liquid (i.e., Fe(l)) under highly reducing conditions ($\leq \Delta$IW-2) following reaction \ref{eq:IW_buffer}. Since $a\text{Fe}=1$ when it precipitates (see Sec. \ref{methods:vapour}), the partial pressure is fixed by $p\text{Fe} = K \cdot a\text{Fe} = K$ (with K the equilibrium constant of the vapourisation reaction $\text{Fe}(l,metal) = \text{Fe}(g)$) for all conditions more reducing than $\sim \Delta$IW $\leq -2$ (dependent on temperature and composition). Thus, the partial pressure of Fe(g) becomes independent of \fOtwo (Fig \ref{fig:melt_vapor_evolution}), and highly reduced atmospheres are dominated by SiO and Mg, even if the melt composition is iron-rich. On the other hand, a minor decrease in $p\text{Fe}$ and $p\text{FeO}$ is observed in highly oxidising atmospheres ($\geq \Delta$IW+3.5), where FeO in the melt is oxidised to Fe$_2$O$_3$ (Eq. \ref{eq:WH_buffer}).

    \item When the melt is oxidising ($\Delta \text{IW} \geq 2$), the atmospheres are dominated by O and O$_2$, and $\log P \sim \log p\text{O}_2 \propto \log f\text{O}_2$. Other species become increasingly diluted.
\end{itemize}

The generalities in the trends of $p$ vs $\Delta \text{IW}$ detailed above are valid for all temperatures tested, but their absolute values (as well as the position of cross-overs between the partial pressures of species) depend on temperature (Fig. \ref{fig:melt_vapor_evolution}). The reason lies in the fact that reaction \ref{eq:f_of_gas_species} depends on absolute \fOtwo, while the partial pressures displayed in Fig. \ref{fig:melt_vapor_evolution} are plotted relative to the IW buffer, the \fOtwo of which increases with temperature.

\subsection{Surface pressure}
\label{results:surface_pressure}

\begin{figure}[t]
    \centering
    \includegraphics[width=\linewidth]{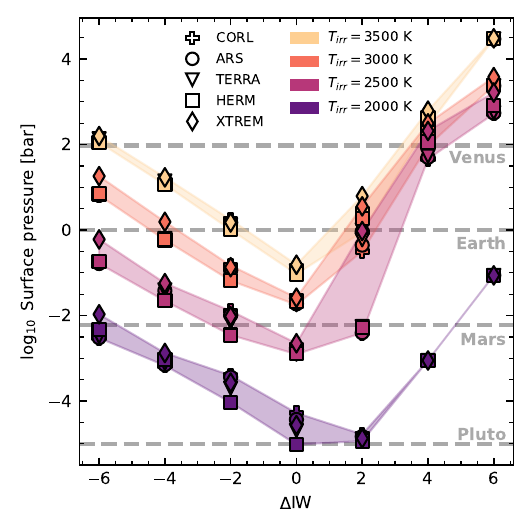}
    \caption{Pressure above a lava ocean as a function of composition, irradiation temperature, and oxygen fugacity (expressed relative to the iron-wüstite buffer, IW), compared to atmospheric pressures of solar-system objects (grey, dashed). Source for pressures: \citet{Catling_Kasting_2017}.}
    \label{fig:surface_pressure}
\end{figure}

Here, we report the pressure at the base of the atmosphere after it has achieved radiative equilibrium (Fig. \ref{fig:surface_pressure}). We find it to be a strong function of irradiation temperature and oxygen fugacity, but not composition. A 500 $^{\circ} C$ increase in temperature roughly increases the atmospheric pressure by a factor $\sim$20, irrespective of composition and \fOtwo. Oxygen fugacity can result in a $\sim$ 5-log unit change in total pressure for a given temperature and composition. We note a particular trend:
\begin{equation}
    \log_{10} P_{\text{surface}} \propto 
    \begin{cases}
    -\frac{1}{2} \log_{10} f_{O_2}  & \Delta \text{IW} \leq 1 \\
    \log_{10} f_{O_2} & \Delta \text{IW} \geq 1 \\
    \end{cases}
\end{equation}
which are a direct result of the systematics described in Sec. \ref{results:vapour_chemistry:fO2}, i.e. from the fact that atmospheres below \dIW{-1} are mainly comprised of metal-bearing gas species (Fe, Mg, SiO), whose partial pressures are proportional to \fOtwo$^{-1/2}$ (Eq. \ref{eq:heterogeneous_equation}), whereas above \dIW{+1}, O$_2$ predominates and hence the total pressure $\sim$ \fOtwo (see also Fig. \ref{fig:melt_vapor_evolution}).  However, oxidised atmospheres do not seem to follow this trend at an irradiation temperature of 2500 K (purple lines); their pressures ramp up suddenly for \dIW{$\ge +2$}, with XTREM transitioning the earliest. This is a result of a greenhouse effect induced by MgO (see Sec. \ref{results:atmospheric_structure}), which affects all oxidised atmospheres (\dIW{$\geq+2$}) and, due to the hotter $T_{BOA}$ of the atmosphere, enhances outgassing.

The minimum in atmospheric pressure is typically attained in the \dIW{$\sim$0-2} range; pressures are $\sim 10^{-4.5}, 10^{-3}, 10^{-2}$ and $10^{-1}$ bar for 2000, 2500, 3000 and 3500 K, respectively. Similar pressures at these temperatures are found by \citet{Miguel:2011, Zilinskas:2022, vanBuchem:2023}. They do not explicitly report \fOtwo, so in order to compare, we estimate their \fOtwo to be $\Delta$IW+3.36, $\Delta$IW+2.53 and $\Delta$IW+1.93 at 2000 K, 2500 K and 3000 K, respectively, supporting the validity of our vaporisation model.\footnote{On the basis that they assumed congruent vaporisation of a Bulk Silicate Earth composition as modelled by \texttt{MAGMA}, it is likely they yield similar results to that given by \cite{visscherfegley2013}. We find excellent agreement between the partial pressures of \citet{visscherfegley2013} and our study when adopting their $\Delta$IW for a given temperature.}

The highest pressures are achieved by ultra-hot (3500 K) and highly oxidised ($\Delta$ IW$>$3.5) atmospheres, reaching up to $\sim 3$ GPa at $\Delta$ IW+6; more moderate temperatures of 2500 K lead to surface pressures of $\sim 500-1000$ bar in case of $\Delta$IW+6. Note that, because all gases are assumed ideal, only qualitative conclusions should be drawn from models at $\Delta$IW+6. Highly reducing atmospheres (\dIW{$\le -2$}) reach pressures of up to 145 bar for 3500 K irradiation, but are more tenuous at lower temperatures, around 0.1-10 bar. This indicates that mineral atmospheres are not necessarily as tenuous as previously thought \citep[i.e., $\sim 3\cdot 10^{-4}, 1.4\cdot 10^{-2}, 2 \cdot 10^{-1}$ and 2 bar at 2000, 2500, 3000 and 3500 K, respectively,][]{visscherfegley2013}, although extreme temperatures and oxygen fugacities are required to achieve pressures beyond 1 bar.


\subsection{Atmospheric structure}
\label{results:atmospheric_structure}

\begin{figure*}
\sidecaption
  \includegraphics[width=12cm]{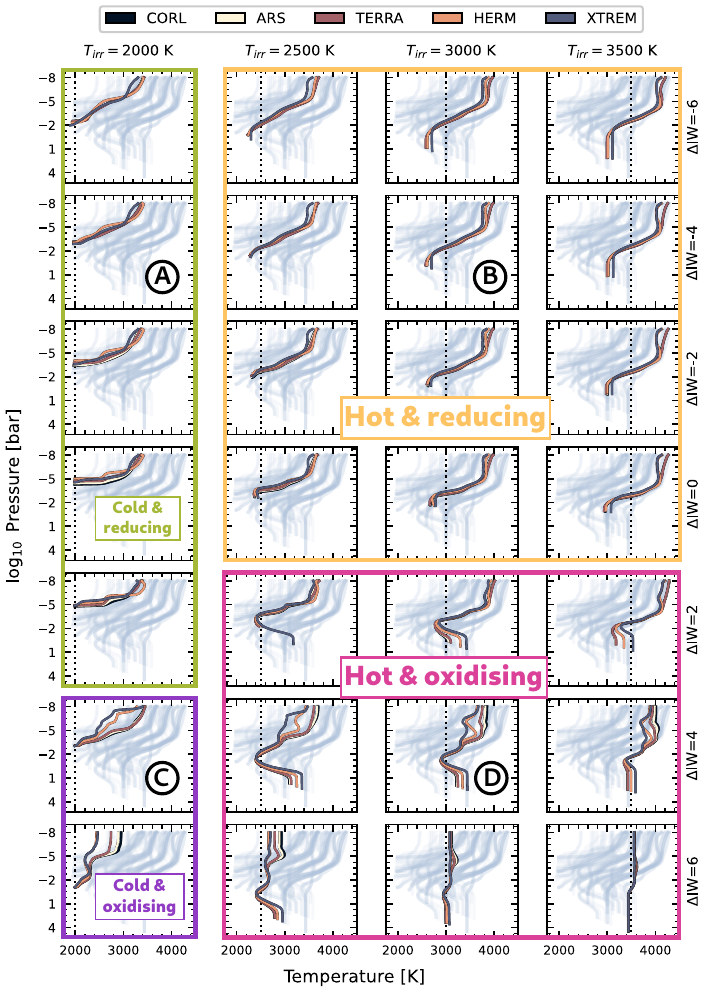}
     \caption{Temperature-pressure profiles of all simulated atmospheres. Each subfigure contains all spectra (light grey), with the ones corresponding to the $T_{\text{irr}}$/$\Delta$IW of the given row/column being highlighted. Color corresponds to composition (Table \ref{tab:archetypes}). The black dotted, vertical lines denote the irradiation temperature of the planet. A-D refers to a selected set of atmospheres to be discussed in more detail in Fig. \ref{fig:lop_representative_atmospheres}.}
     \label{fig:p_T_profiles}
\end{figure*}

\begin{figure*}[p]
    \centering
    \includegraphics[width=0.95\textwidth]{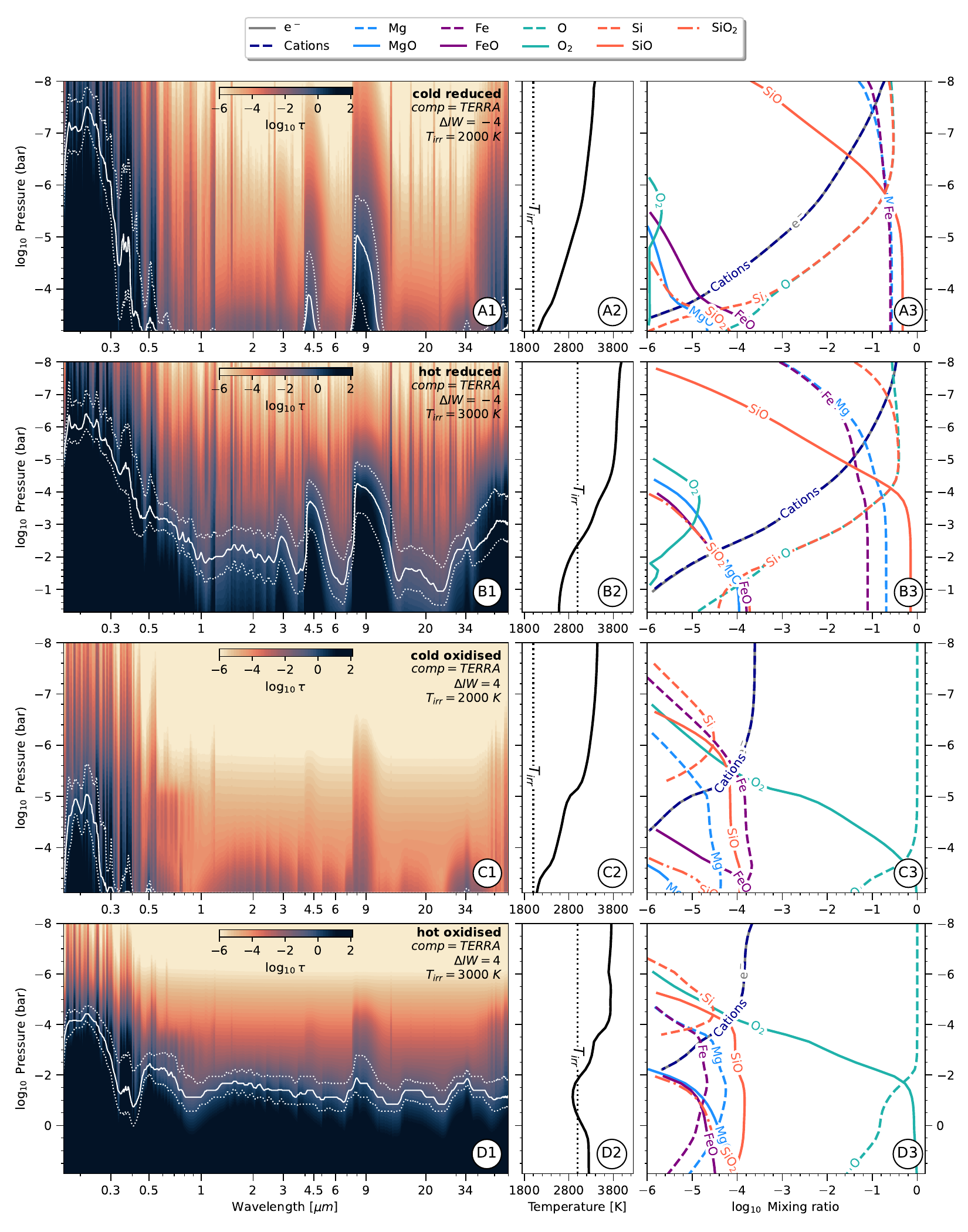}
    \caption{Atmospheric structure of representative combinations of $\Delta$IW and $T_{\text{irr}}$, selected from Fig. \ref{fig:p_T_profiles}, A-D. The left column displays the optical depth in every atmospheric layer (color gradient), as well as an indication of the photospere (white). The latter is defined where the atmosphere has absorbed 16\% (upper dotted), 50\% (solid) and 84\% (lower dotted) of all incoming light. The center displays the atmospheric pressure-temperature profile, with the irradiation temperature $T_{\text{irr}}$ highlighted as vertical dotted line. On the right, the (volume) mixing ratios of major chemical species are shown.}
    \label{fig:lop_representative_atmospheres}
\end{figure*}

The 140 simulated atmospheres can be broadly grouped by temperature and \fOtwo regime based on their speciation, optical thickness, and pressure-temperature structure. Composition was found to exert only minor influences on most characteristics, so it is ignored in the classification. The resulting classes are labelled "cold \& reducing", "cold \& oxidising", "hot \& reducing" and "hot \& oxidising", based on their $P$-$T$-profiles shown in Fig. \ref{fig:p_T_profiles} and their speciation \& physical characteristics (e.g. Fig. \ref{fig:lop_representative_atmospheres}). From each class, we select a representative endmember in fugacity and temperature, which are labeled A to D in Fig. \ref{fig:p_T_profiles}; the chosen melt composition is TERRA. The atmospheric characteristics of these representative endmembers are displayed in Fig. \ref{fig:lop_representative_atmospheres} and will be discussed in the following. We note, however, that this is only a broad classification, mostly based on our discrete sampling in oxygen fugacity, temperature and composition space; in reality, all these cases transition into each other continuously.

In order to discuss the radiation physics of these atmospheres, we introduce the concept of optical depth or, as outlined here, its change along a single path of length $z$ taken by a lightray as
\begin{equation}
    \label{eq:optical_depth}
    \Delta \tau = \kappa \rho z
\end{equation}
where $\kappa$ is the opacity in units of area per mass, e.g. $cm^2 g^{-1}$ (see Section \ref{methods:opacities}), $\rho$ is the mass density and $z$ is the thickness of the absorbing layer. The transmissivity $t$ is then defined by the following function (the Beer-Lambert-law):
\begin{equation}
    t = \frac{I_{\text{out}}}{I_{\text{in}}} = e^{-\Delta \tau}
\end{equation}
where $I_{\text{in/out}}$ are the incoming and outgoing intensity of the attenuated ray. These simple considerations apply to every layer of the atmosphere. We make use of $\tau$ in Fig. \ref{fig:lop_representative_atmospheres} to demonstrate how the radiation interacts with the atmospheric chemistry. Additionally, we will define the integrated transmissivity:
\begin{equation}
    \label{eq:integrated_transmissivity}
    t_k = \prod_{i=0}^k t_i
\end{equation}
where $t_k$ is the integrated transmissivity of layer $k$ and $t_i$ is the transmissivity of atmospheric layer $i$, with $i=0$ being the top of the atmosphere. This describes which fraction of a radiation beam coming from space has already been absorbed by the time it reaches layer $k$ (or, by invoking time symmetry, how strongly the outgoing radiation from layer $k$ has been attenuated). Setting $t_k=50$\% provides us with a tentative definition of the photosphere, displayed by the white ribbons in Fig. \ref{fig:lop_representative_atmospheres}, A1-D1.

We further define the (volumetric) mixing ratio of a species $i$ in a gas as \citep{Catling_Kasting_2017}:
\begin{equation}
    v_i = \frac{N_i}{N}
\end{equation}
where $N_i$ is the number of moles of species $i$ in a gas, and $N$ is the total mole number of the gas.

\subsubsection{Cold and reducing}
\label{results:atmospheric_structure:cold_reducing}

Outgassed atmospheres of `cold' ($\sim 2000$ K), intermediate to reducing (below $\Delta$IW+2) lava planets are predominantly composed of SiO, Si, MgO, Fe (if available) and monoatomic O (Fig. \ref{fig:lop_representative_atmospheres}, A3); the latter being as abundant as Si(g) throughout the atmosphere as a result of dissociation of SiO, the main carrier of oxygen. Free electrons from thermal ionisation become abundant towards the upper atmosphere, making it a potential conductor, albeit the atmosphere remains electrically neutral due to the remnant cations, which are predominately singly ionised (cf. Fig. \ref{fig:lop_representative_atmospheres} A3 and Table \ref{tab:fastchem_reactions}).

The relative stability of SiO throughout the atmospheric column has a strong effect on the pressure-temperature profile: since it is an effective absorber in the UV, the atmosphere becomes almost fully opaque at short wavelengths ($\le 0.3 \mu$m; Fig. \ref{fig:lop_representative_atmospheres}, A1). The absorbed heat has to be re-emitted in the infrared, which is mostly transparent except for the strong bands of SiO at 4.5 and 9 {\textmu}m. This requires the upper layers to heat up in order to achieve radiative balance by emission through the only available lines, and as a result, the atmospheric profile develops a thermal inversion \citep{Malik:2019, Ghandi_2019_TiOinversion}. In this class, the inversion extends throughout the entire atmospheric column (Fig. \ref{fig:lop_representative_atmospheres}, A2). Similar inversions were also found by \citet{Ito:2015} and \citet{Zilinskas:2022}.

Atmospheric profiles of this class are largely independent of melt composition. This can be explained by SiO being the most important (and abundant) absorber/emitter in all reducing atmospheres (see Sec. \ref{results:vapour_chemistry}), with Si, Mg, Fe and O gases being consigned to play minor roles. It is worth mentioning that the emission of SiO in the infrared blocks important spectral bands (located at 9 {\textmu}m) for the characterisation of silicate surfaces \citep{Hu_2012, Fortin:2022}. Otherwise, the atmosphere remains transparent throughout the infrared (see Fig. \ref{fig:lop_representative_atmospheres} A1). The relative transparency of these atmospheres mean that the melt temperature closely approximates the irradiation temperature.

\subsubsection{Hot and reducing}
\label{results:atmospheric_structure:hot_reducing}

Atmospheres of hot ($T_{\text{irr}} \geq 2500$ K) and reducing ($\Delta$IW $\leq 0$) planets are similar to their cooler counterparts (abundant Mg, Fe, Si and O), but all have uniformly higher partial pressures, including SiO(g), due to the higher temperatures \citep{Miguel:2011, Wolf:2022} (Fig. \ref{fig:lop_representative_atmospheres}, B3). However, high temperatures also cause SiO to dissociate more readily into weaker UV absorbers Si and O, which makes the upper layers of the atmosphere ($p \leq 10^{-5}$ bar) slightly more transparent than in cooler, equally reducing planets (i.e., the photosphere is situated at $\sim 10^{-6}$ bar in the hot case vs. $\sim 10^{-7}$ bar in the cold case, Fig. \ref{fig:lop_representative_atmospheres}, A1 vs. B1). Consequently, the capacity to absorb UV radiation is reduced, resulting in the development of (nearly) isothermal parts in the upper atmospheres; the isotherms in XTREM and HERM are more pronounced due to their even lower SiO content. Similar to the cold and reduced case, free electrons and cations are abundant in this layer.

Closer to the surface, at pressures higher than $\sim$ 10$^{-4}$ bar, SiO becomes the predominant species, again leading to a strong thermal inversion and hence to a temperature decrease towards the surface. Contrary to the cold \& reducing case, this inversion is confined to intermediate atmospheric layers, and it cools the bottom layers of the atmospheres below the irradiation temperature of the planet (c.f. Fig. \ref{fig:p_T_profiles}).

Underneath the inversion, at pressures $\geq 0.1$ bar, the higher pressures and cooler temperatures stabilise molecules such as MgO, FeO and TiO. In particular, MgO is a strong, grey absorber in the infrared, and in combination with SiO it traps light in essentially all wavelength ranges, leading to optically thick lower layers \citep[see also][]{Zilinskas:2022}. Diffusive transport of radiation is initiated, as evidenced by the (approximately) isothermal $P$-$T$-profiles in the lower atmosphere (Fig \ref{fig:p_T_profiles} upper right and Fig. \ref{fig:lop_representative_atmospheres} B2), which becomes more pronounced in hotter, more reducing planets.

\subsubsection{Cold and oxidising}
\label{results:atmospheric_structure:cold_oxidising}

Cold ($\sim$ 2000 K) and oxidising ($\geq \Delta$IW+4) lava ocean planets produce atmospheres mainly composed of monatomic O. Diatomic O$_2$ is also present, but dissociates in the layers above the lava ocean as temperatures increase (Fig. \ref{fig:lop_representative_atmospheres} C3). Other species are scarce (with mixing ratios $\leq 10^{-4}$), though chief among them are Fe, SiO and Mg, whereas Si is less abundant compared to the reducing class of atmospheres (due to the homogeneous reaction $Si(g) + \frac{1}{2}O_2(g) \rightarrow SiO(g)$ being favoured). Additionally, FeO reaches relatively high mixing ratios in the lower atmospheres (with the potential to exceed SiO), but dissociates in the upper layers into Fe(g) and O(g). Free electrons and ions increase in abundance in the upper atmosphere, but the ionisation degree remains lower than in a reducing atmosphere.

Since SiO is no longer present in high mixing ratios ($\leq 10^{-4}$), its role as a UV absorber is diminished, and only becomes relevant at relatively high pressures (above $10^{-6}$ bar) in the lower atmosphere. Further, its number density remains low enough such that the infrared bands - which are weaker than the UV bands - cause only minor attenuation of the in- and outgoing radiation and hence remain negligible (Fig. \ref{fig:lop_representative_atmospheres} C1). Thermal inversions are still present, but weaken as the melt-vapour system moves towards $\Delta$IW+6, a result of lower $p_{SiO}$. Due to the comparatively low temperature of the planet ($\sim 2000$ K), the atmosphere never becomes thick enough for SiO, MgO and TiO to act as efficient absorbers, rendering it relatively transparent throughout the entire infrared wavelength range, implying that the surface is observable from space. The transition from relatively transparent to fully opaque at the 9 {\textmu}m feature - important for surface characterisation - occurs between $\Delta$IW+2 and $\Delta$IW+4. 

Interestingly, this atmospheric class is sensitive to composition (see Fig. \ref{fig:p_T_profiles}, lower left). In particular, the upper atmospheres of the iron-free archetypes HERM and XTREM remain cooler than atmospheres with iron vapour (but are still in excess of the irradiation temperature). There is also a correlation between the iron content of an atmosphere and the temperature of the upper layers under highly oxidising conditions ($>\Delta$IW+3.5). As SiO(g) has significantly lower mixing ratios than in the reducing cases, the relative opacity of Fe(g) becomes more important in shaping the upper atmosphere, thus imparting the difference in melt FeO content to the $P$-$T$-profile of the atmosphere. However, the majority of differences between HERM \& XTREM vs. CORL-ARS-TERRA lies within the lower SiO$_2$(l) abundance of the former group, which leads to lower mixing ratios of SiO(g) and therefore weaker inversions (Fig. \ref{fig:p_T_profiles}).

\subsubsection{Hot and oxidising}
\label{results:atmospheric_structure:hot_oxidising}

Hot and oxidising atmospheres extend to significantly higher pressures than their cooler counterparts (in most cases > 1 bar, cf. Fig. \ref{fig:surface_pressure}), but are similar in chemistry (dominated by O$_2$ and O, with low $p_{SiO}$), transparency (the photosphere is located at comparatively high pressures, $10^{-2}$ bar) and the weakening thermal inversions with increasing \fOtwo. The atmospheric inversions almost disappear at $\Delta$IW+6 and $T_{\text{irr}} \geq 3000$ K, becoming near isothermal. This is due to the dilution of SiO by O and O$_2$, as well as suppressed $p$SiO in absolute terms.

Further down the atmospheric column, the same effect leaves MgO(g) as the most efficient absorber in the atmosphere; its opacities span over nearly all wavelengths except for the shortwave bands ($\le 400$ nm) and are intrinsically higher than for SiO (see Fig. \ref{fig:molecular_opacities}); hence, since MgO(g) is present in similar mixing ratios as is SiO at lower altitudes, it causes the atmosphere to become an efficient infrared emitter, leading to a flattened photosphere (Fig. \ref{fig:lop_representative_atmospheres} D1). Below the photosphere, the atmosphere is optically thick and radiation must be transferred diffusively. 

Another striking feature that is not seen in other classes is a distinct cold trap, expressed in all but the hottest and most oxidising cases. It occurs in the mid-atmosphere just below the photosphere (0.001 -- 1 bar) (Fig. \ref{fig:lop_representative_atmospheres}, D2), and arises from a translucent optical window spanning the range 0.3 to 0.5 {\textmu}m (Fig. \ref{fig:lop_representative_atmospheres}, D1). The window enables a substantial portion of stellar radiation to penetrate deeper atmospheric layers, heating them from below. Since the atmosphere is optically thick in IR at the pressure level of the window, it lacks the ability to re-emit to space, resulting in radiation entrapment which ultimately causes a greenhouse effect. The window is a consequence of a low SiO mixing ratio; in its wavelength range, there is a particular drop in the opacity continuum (cf. Fig. \ref{fig:molecular_opacities}) unless either SiO or TiO appear in large mixing rations, both of which have low partial pressures in oxidising atmospheres.

Similar to their cooler counterparts, hot and oxidising atmospheres show some response to the iron content in their upper atmospheres, for the same reason. Only for ultra-hot and oxidising cases ($T_{\text{irr}} \geq 3000$, $\Delta$IW+6) does this effect nearly vanish together with the thermal inversion due to the absolute dominance of oxygen. 


\begin{figure*}[p]
    \centering
    \includegraphics[width=0.98\textwidth]{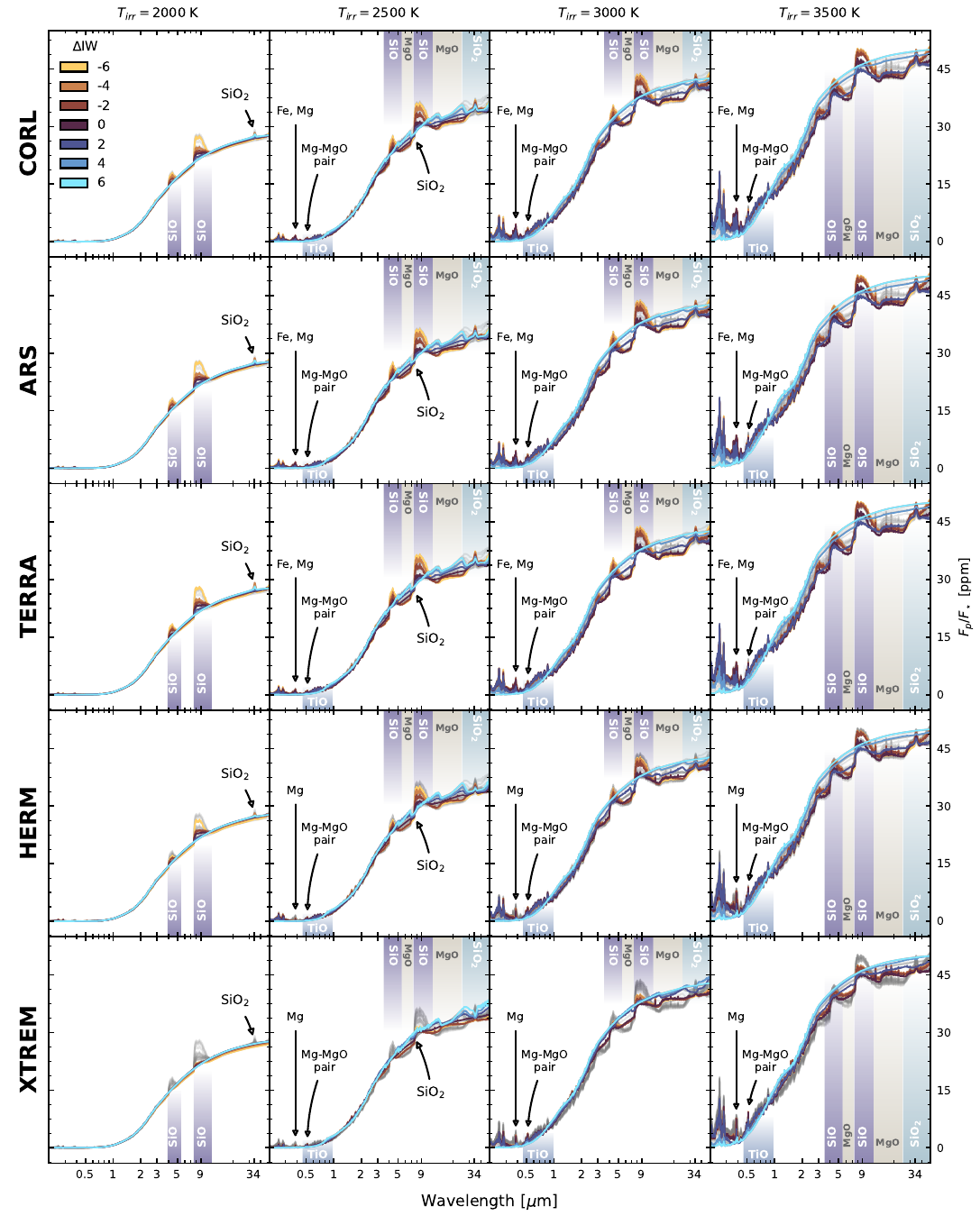}
    \caption{Spectra of all 140 simulations of this study. Each column contains all spectra of equal irradiation temperature, and each row constant composition. In each box, all spectra of constant irradiation temperature are plotted (grey), but only the ones corresponding to the composition indicated in the row are coloured. The colour-coding is made according to their oxygen fugacity (yellow most reducing, brown intermediate, light blue most oxidising). Species responsible for important spectral features are indicated by the pointers, important wavebands by coloured patches.}
    \label{fig:spectra_all}
\end{figure*}

\subsection{Spectra}
\label{results:spectra}

Due to their proximity to the star and expected tidal locking, atmospheres of LOPs are most readily studied during secondary occultations. This involves capturing a time series of spectra during the brief moments when the planet transits behind its host star. In the intervals just before and after the eclipse, the planet's dayside is revealed, resulting in a combined flux of both the planet ($F_p$) and the star ($F_s$). The observed brightness difference is a function of the flux ratio between the planet and planet+star combination, which can be expressed as \citep[e.g.,][]{seager2010}

\begin{equation}
    \frac{F_p}{F_s} = \frac{\varepsilon_p{R_p}^2}{\varepsilon_s{R_s}^2}
    \label{eq:flux_ratio}
\end{equation}

where $R_p$ and $R_s$ are the radii of the planet and star, respectively, and $\varepsilon_p$ and $\varepsilon_s$ denote their respective spectral exitance (i.e., the energy flux emitted per wavelength and surface area, given here in erg s$^{-2}$ cm$^{-3}$). This unique scenario facilitates the separation of the planet's spectral information from that of its host star. Observation of LOPs during the secondary eclipse is favourable due to the fact that \textit{i.)} their atmospheres are expected to have higher molecular weights ($M$) than Hot Jupiters or Sub-Neptunes, which renders transit spectroscopy more challenging because of their low scale height, $H$;

\begin{equation}
    H = \frac{RT}{Mg}
    \label{eq:scale_height}
\end{equation}

where $R$ is the gas constant and $g$ the acceleration due to gravity, and \textit{ii.)} the anticipated condensation of vapour toward the terminator, leaving little to no atmosphere along the planetary limb to be studied \citep[e.g.][]{Nguyen:2021}. Throughout this study, we may interchangeably refer to the depth of the secondary eclipse as secondary eclipse depth, occultation depth or planet-to-star flux ratio. However, using the secondary occultation depth means that observables unrelated to the spectral exitance of the planet, such as the stellar emission and the planetary radius, are convoluted with the result (see Eq. \ref{eq:flux_ratio}). However, the spectral exitance of the modelled atmospheres is expected to vary negligibly with the characteristics of the planet, and the occultation depth of a larger planet with identical $T_{\text{irr}}$, \fOtwo, composition and host-star type to the simulations shown here (1 $M_\oplus$, 1 $R_\oplus$) should scale-up by $R_p^2$, where $R_p$ is in $R_\oplus$.

\subsubsection{General features}
\label{results:spectra:general_features}

Secondary occultation depths in Fig. \ref{fig:spectra_all} reveal that mineral atmospheres show features over a vast range of spectral domains. It should be borne in mind that, although the flux ratio increases as a function of wavelength, the absolute flux declines beyond the black body emission peak, making features beyond $\sim$ 20 {\textmu}m increasingly challenging to detect. Many spectral features, largely derived from molecular species (MgO, SiO and SiO$_2$) are located in the mid-infrared (MIR, loosely defined here as 3-25 {\textmu}m). TiO also has some strong bands that appear similar in shape and strength to the SiO features (cf. Fig. \ref{fig:molecular_opacities}), however, owing to its low atmospheric mixing ratios they never become relevant. All these species should appear in atmospheres derived from any melt composition, as both Si and particularly Mg are lithophile. Moreover, this region is also accessible to modern or planned missions such as the JWST or Ariel.

Further, we find a plethora of lines in the ultraviolet (UV), visible (VIS) and near-infrared (NIR) up to $\sim 1$ {\textmu}m. A considerable contribution comes from neutral atoms, which are particularly abundant at high altitudes in mineral atmospheres (see Fig. \ref{fig:lop_representative_atmospheres}). For the most part, these lines are produced by Mg and Fe (300 -- 520 nm), with some contribution from MgO (480 nm). The UV is dominated by emission from SiO, where it has its strongest feature (see Fig. \ref{fig:molecular_opacities}); but also the strong doublet line of Mg (280 nm) is located there, as are some lines of Fe. In the visible and very near infrared (500 nm – 1 {\textmu}m), a mixture of MgO and TiO lines are present. The latter has stronger opacities per unit mass (Fig. \ref{fig:molecular_opacities}) but MgO has higher mixing ratios and therefore contributes more than does TiO. Some weaker lines of Fe might also contribute in the NIR, but their opacities are lower than those of the aforementioned molecular species and therefore do not generate observable features at the resolution used to generate spectra by our model. 

\subsubsection{Effect of composition}
\label{results:spectra:effect_of_comp}

Few substantial differences in the flux ratio of the MIR spectra can be ascribed to the composition of the underlying melt, regardless of \fOtwo or temperature (Fig. \ref{fig:spectra_all}). Compositions TERRA, ARS and CORL are nearly identical, even though the iron content varies significantly between them. HERM shows some slight reduction in the strength of the SiO features, yet follows the other compositions closely despite its considerably lower SiO$_2$/MgO ratio ($\sim 0.863$ instead of $\sim1.16$, the value for TERRA, CORL and ARS). Only the extremely reduced planetary archetype XTREM is distinct from its peers, as it has weaker SiO emission at both 4.5 and 9 {\textmu}m owing to its low SiO$_2$ content. This observation can be reconciled with the predictions from the vaporisation model, Eq. \ref{eq:heterogeneous_equation}, that $pSiO$ should rely on the activity (and therefore concentration) of SiO$_2$ in the melt.

\begin{figure}[t]
    \centering
    \includegraphics[width=\linewidth]{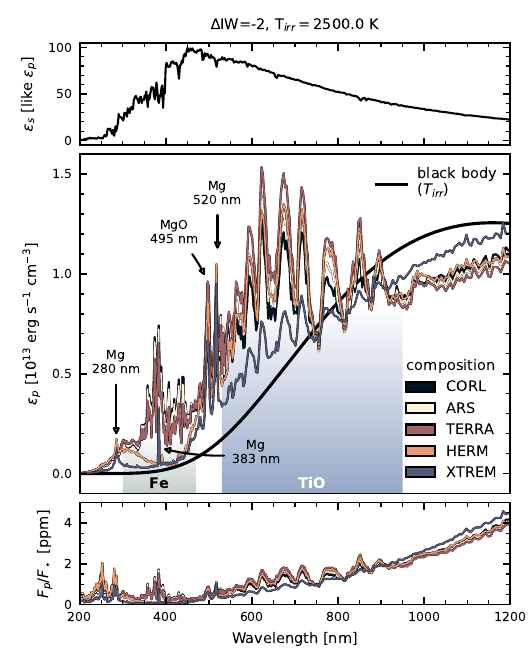}
    \caption{Effect of composition on the spectrum of a mineral atmosphere at constant \fOtwo relative to the IW buffer ($\Delta$IW-2) and irradiation temperature (2500~K) for an Earth-sized planet. Shown are the stellar spectrum (top), the spectrum of planetary emission (center, coloured) and the secondary occultation depth $F_p/F_\star$, the ratio between the two (bottom, colours). Flux ratios for outgassed atmospheres may overlap, as evident for the CORL and ARS compositions. Only the ultra-SiO$_2$-poor composition XTREM ist slightly distinct by showing less emission in the TiO band. Both iron-free compositions (HERM and XTREM) do not express iron lines (red field).}
    \label{fig:spectrum_composition}
\end{figure}

In the UVIS, the effect of composition under constant \fOtwo ($\Delta$IW = -2) and temperature ($T_{\text{irr}}$ = 2500 K) on the emission spectrum are shown in Fig. \ref{fig:spectrum_composition}; other $T_{\text{irr}}$ would induce similar effects with stronger lines (compare Fig. \ref{fig:spectra_all}), while the effect of \fOtwo is discussed in Section \ref{results:spectra:effect_of_fO2}.
One of the major compositional controls is the emission from 330 to 460 nm by Fe(g), which differs markedly between the two groups of planetary mantles that contain FeO (CORL, ARS, TERRA) and those that do not (HERM, XTREM). We also notice a consistent, albeit subtle decrease of the Fe line intensity with increasing FeO(l) (CORL has highest $p$Fe abundance, TERRA the lowest). 

The other species reacting to melt composition is TiO via its broadband emission from 0.5 -- 1 {\textmu}m, which varies by a factor $\sim$ 3 from XTREM to TERRA. However, while the variation within the archetypes CORL, ARS, TERRA and HERM corresponds to the abundance of TiO$_2$ in the melt (Table \ref{tab:archetypes}), XTREM shows an unusually low TiO(g) emission despite its higher abundance of TiO$_2$. This suppression of emission is probably not realistic\footnote{This is a result of the low activity coefficient: $\gamma_{\text{XTREM, TiO}_2} \sim 0.015$ vs $\gamma_{\text{TERRA, TiO}_2} \sim 0.17$. This results in an order of magnitude lower partial pressures for TiO(g) in XTREM, thus lowering emission. The small $\gamma$ is probably due to a breakdown of the activity model in \texttt{MAGMA}; with the results of \citet{borisov_2020_tio2} we find $\gamma_{\text{TiO}_2} \sim 1.1$ at the respective melt temperature, although this still involves an extrapolation both in composition and temperature.}, but serves to illustrate the effect of changing TiO$_2$(l). The increased flux of XTREM in the MgO band beyond 1 {\textmu}m (Fig. \ref{fig:spectrum_composition}) is a result of increased heating of the lower atmosphere due to lower TiO(g). 

\subsubsection{Effect of oxygen fugacity}
\label{results:spectra:effect_of_fO2}

Oxygen fugacity has a considerable impact on the MIR flux ratio for the planets on our grid (Fig. \ref{fig:spectra_all}). Most obviously, the intensities of the SiO features at 4.5 {\textmu}m and 9 {\textmu}m are more strongly expressed at reducing conditions ($\Delta$IW $\leq 0$), whereas they vanish under oxidising conditions ($\Delta$IW $\geq 2$), to the point where atmospheres at $\Delta$ IW $\geq 4$ mimic a black body. Factors controlling the intensity of SiO emission features are: 
\begin{itemize}
    \item increased opacity through increased mixing ratios of SiO(g), where $p$SiO increases as a function of \fOtwo$^{-0.5}$ (see Sec. \ref{methods:vapour} and \ref{results:vapour_chemistry})
    \item increased opacity through pressure broadening: the total pressures are higher for reducing- ($\Delta$IW $\leq 0$) with respect to intermediate (0 $ \leq \Delta$IW $ \leq 2$) atmospheres since the partial pressures of the monatomic gases of other major elements (Fe, Mg) also depend on \fOtwo$^{-0.5}$ (see Sec. \ref{results:surface_pressure}).
    \item thermal inversions, which are more prevalent in reducing- and intermediate atmospheres, but weaker in oxidising atmospheres (see Sec. \ref{results:atmospheric_structure}); it is this property that produces SiO emission features in the first place. 
\end{itemize}
The corollary is that atmospheres produced in equilibrium with a more reduced silicate liquid produce stronger SiO(g) emission, while atmospheres derived from oxidised magmas ($> \Delta$IW+2) have barely distinguishable SiO emission features relative to the black body baseline (Fig. \ref{fig:spectra_all}).

The second most striking feature in the MIR is the difference in MgO background emission (blue fields, Fig. \ref{fig:spectra_all}) between oxidising (blue) and reducing (yellow) cases. Because $p$MgO is independent of \fOtwo, it is diluted  by SiO(g), Mg(g) and Fe(g) (if present) in reducing atmospheres. Therefore, it can achieve high mixing ratios only at higher pressures, which pushes the photosphere in the MgO-bands to lower layers cooled by TiO(g) emission, thus decreasing emission in MgO bands (see Appendix \ref{apx:TiO}). This results in enhanced contrasts between the photosphere (dominated by the MgO background) and the SiO features at 4.5 and 9 {\textmu}m, from essentially nil at $\Delta$IW+6 and +4 to $\sim 10$ ppm (Earth-sized planet) at $\Delta$IW-6. Larger and hotter planets should express stronger absolute contrasts; a 2 $R_\oplus$, 2500 K planet might reach $\sim 40$ppm contrast, a similarly sized 3000 K planet could achieve $\sim 60$ ppm (cf. eq. \ref{eq:flux_ratio}). The linearity in the spectral contrast over this wavelength range holds promise for the deduction of \fOtwo in such atmospheres; a more detailed discussion can be found in Sec. \ref{discussion:degeneracy_between_meltcomp_and_fO2}.

Previously, it has been shown that, in addition to the SiO(g) emission feature, SiO$_2$(g) can be detected from an absorption feature at 7 {\textmu}m in MIR-observations of LOPs with JWST \citep{Zilinskas:2022} and can carry information on the atmospheric \fOtwo \citep{Wolf:2022}. We find the SiO$_2$ 7 {\textmu}m absorption feature to typically exist in oxidising atmospheres ($\Delta$IW +2 to +4) of 2500 -- 3000 K, which corresponds to the \fOtwo expected in \citet{Zilinskas:2022}. In more reducing systems it is expressed as an emission feature that is roughly independent of \fOtwo (as expected from Eq. \ref{eq:f_of_gas_species}) and thus could serve as a baseline to compare the SiO(g) feature to, in order to derive \fOtwo. However, it is of invariably weaker intensity than the adjacent MgO and SiO features, making it comparatively difficult to quantify.

\begin{figure}[t]
    \centering
    \includegraphics[width=\linewidth]{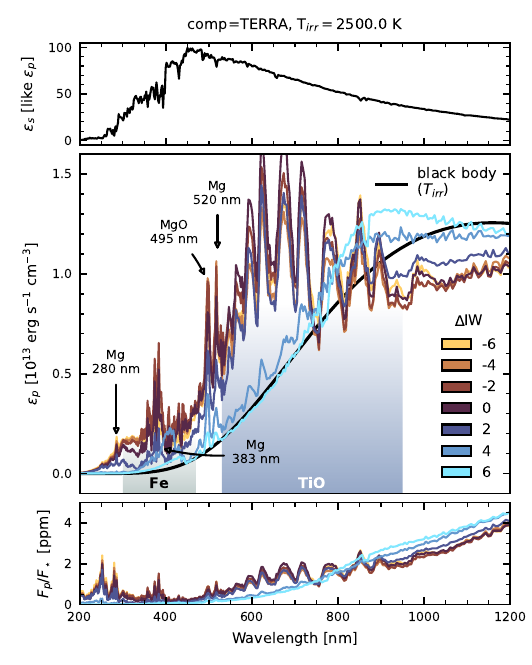}
    \caption{Effect of \fOtwo on the spectrum of a mineral atmosphere at constant melt composition (TERRA) and irradiation temperature (2500~K) for an Earth-sized planet. Shown are the stellar spectrum (top), the spectrum of planetary emission (center, coloured) and the secondary occultation depth $F_p/F_\star$, the ratio between the two (bottom, colours). The emission from metallic species (Fe, Mg, MgO, TiO) in the visible distorts the overall shape of the spectral energy distribution (SED) compared to a pure black body. This effect is pronounced in reducing atmospheres, whereas oxidised ones follow a black body more closely except for a handful of strong lines.}
    \label{fig:spectrum_fO2}
\end{figure}

The UVIS/NIR spectrum is also affected by \fOtwo (Fig. \ref{fig:spectrum_fO2}). However, even though the planet strongly radiates via these lines, the secondary eclipse depth is low ($\leq 2$ ppm for an Earth-sized planet) due to the brightness of the star in this wavelength range, though it increases with temperature (Fig. \ref{fig:spectra_all}).
As highlighted in Fig. \ref{fig:spectrum_fO2}, the iron lines between 330 to 460 nm respond to changes in \fOtwo. This could make them potential tools to trace \fOtwo, however, FeO contents can vary greatly owing to its extraction into planetary cores, and iron line intensity in this range should be coupled with other \fOtwo-sensitive spectral feature(s). Moreover, the intensity of iron lines is not a linear function of log\fOtwo, owing to the saturation in Fe metal at low \fOtwo (see Sec. \ref{results:vapour_chemistry}), meaning maximum emission occurs at $\Delta$IW-2 (shifting towards $\Delta$IW+0 for higher temperatures, c.f. Fig. \ref{fig:spectra_all}). The partial pressure of Fe(g) decreases with \fOtwo$^{-0.5}$, such that oxidising atmospheres ($\geq \Delta$IW+2) have no remarkable iron lines and instead an emission ``bump" is observed, where hotter, lower layers radiate through the optical window formed by the paucity of SiO(g) (see Sec. \ref{results:atmospheric_structure}). 

Another notable difference is caused by the presence of TiO(g). Towards the visible wavelength range, we find that its emission ($\sim 0.45-1$ {\textmu}m) is suppressed for highly oxidising atmospheres ($\Delta \text{IW} \geq 4$), due to its low mixing ratios, a result of lower partial pressures (see Eq. \ref{eq:f_of_gas_species}) and its dilution by O$_2$ in the atmosphere. Reducing conditions, on the other hand, produce strong TiO features (except for XTREM, see Sec. \ref{results:spectra:effect_of_comp}); however the variation in intensity as a function of \fOtwo is not linear.

Lastly, the Mg and MgO peaks at 495 and 520 nm, respectively, grow in intensity as conditions become more reducing. Their ratio changes: in reducing atmospheres, the Mg(g) feature is more pronounced, whereas MgO(g) shows the higher peak under oxidising conditions. However, we note that the resolution of our opacities (0.01 cm$^{-1}$) is not sufficient to fully capture the shape of such narrow, highly peaked lines.


\section{Discussion}
\label{sec:discussion}

\subsection{Alkali metals}
\label{discussion:alkali}

\begin{figure}[!t]
    \centering
    \includegraphics[width=\linewidth]{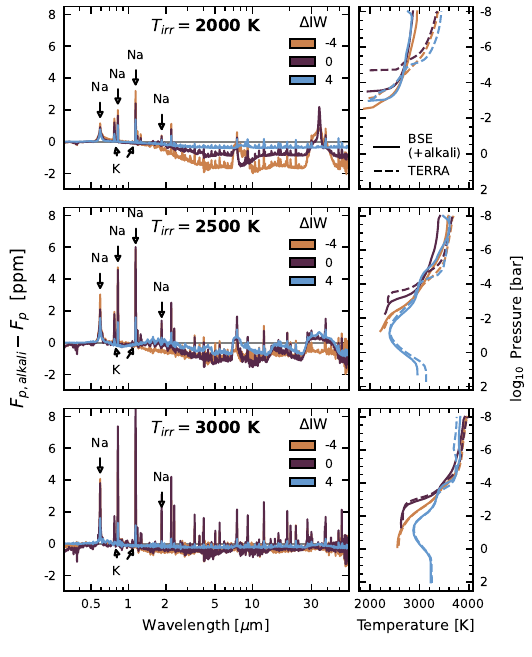}
    \caption{Influence of alkali metals (Na, K) on structure and spectrum of mineral atmospheres of an Earth-sized planet. On the left hand side, the difference in the planet-to-star flux ratio between an atmosphere with and without alkali metals is shown. The composition is TERRA (similar to bulk silicate Earth, BSE) with 0.36 wt. \% Na$_2$O and 0.029 wt. \% K$_2$O added \citep{mcdonough1995}.}
    \label{fig:alkali_test}
\end{figure}

We deliberately studied alkali-free systems, even though these elements have been theorised to occur in abundance in lava planet atmospheres \citep{Fegley:1987, Schaefer:2004, Leger:2011, Ito:2015, Zilinskas:2022}. However, their volatility during planet-forming processes \citep{larimer1967,sossi2019,gellissen2019} means that their abundances in the planet cannot be estimated reliably from those in the star, in contrast to more refractory elements such as Si, Mg, Fe, Al, Ca and Ti. If Na$_2$O and K$_2$O were present in the melt, the resulting atmospheres would become rich in their respective metal-bearing gases: Na, K, NaO and KO \citep{Fegley:1987, Schaefer:2004, Zilinskas:2022, Wolf:2022, vanBuchem:2023}. These gases would have high mixing ratios, particularly for low temperature ($\sim$ 1500 K) planets with tenuous atmospheres, and would absorb in the visible at their respective doublet lines around 589 and 769 nm.

In order to quantitatively determine the effect of alkalis on the compositions, $P$-$T$ profiles and emission spectra of mineral atmospheres, 
the BSE composition from \citet{mcdonough1995} (nearly identical to TERRA, Table \ref{tab:archetypes}, but containing 0.36 wt\%  Na$_2$O and 0.029 wt\% K$_2$O) were used and modelled at $\Delta$IW-4, IW and $\Delta$IW+4 (Fig. \ref{fig:alkali_test}). As a consequence, the planet exhibits strong but narrow emission lines, particularly of Na, in the visible \& NIR (Fig. \ref{fig:alkali_test}, left). This causes a slight cooling effect due to the increased emissivity of the upper atmosphere \citep[see also][]{Zilinskas:2022}; its magnitude depends on the temperature and \fOtwo of the planet. It is most pronounced for cold planets ($T_{\text{irr}}\sim 2000$ K, all redox states) and the $\Delta$IW = 0 case at 2500 K. Monatomic Na becomes the dominant gas in such atmospheres (Fig. \ref{fig:BSE_fugacity_series} because $p$Na $\propto fO_2^{-0.25}$, though it is overwhelmed by SiO at lower \fOtwo, and, at higher \fOtwo, by O$_2$). This also explains their higher surface pressures relative to the alkali-free cases (Fig. \ref{fig:alkali_test}, right panel), consistent with the findings of \cite{Fegley:1987, Miguel:2011, Ito:2015, Zilinskas:2022}. Hence, alkalis contribute to an increase in the total pressures of all atmospheres below $T_{\text{irr}} \sim 2500$ K. For planets with $T_{\text{irr}} \geq 2500$ K, the overall impact on MIR spectra remains limited, and hence does not jeopardise the conclusions drawn earlier.

The volatility of sodium and potassium in the context of lava ocean evaporation means that a larger mass fraction of their total budget resides in the atmosphere relative to the other rock-forming elements considered here. Thus, continuous removal of atmospheric mass through condensation or escape to space would deplete Na and K in the residual magma more rapidly than for other elements, leaving behind the more refractory species \citep[cf.][]{schaefer2009chemistry,Kite:2016, Nguyen:2021, erkaev2023}. This, together with the weaker dependence of $p$(Na,K) on \fOtwo$^{-0.25}$ compared to other species like SiO(g) or Fe(g), means spectral features of alkali metals are, alone, poor indicators of \fOtwo.

\subsection{Degeneracy between melt SiO$_2$/MgO ratio and \fOtwo}
\label{discussion:degeneracy_between_meltcomp_and_fO2}

\begin{figure}[t]
    \centering
    \includegraphics[width=\linewidth]{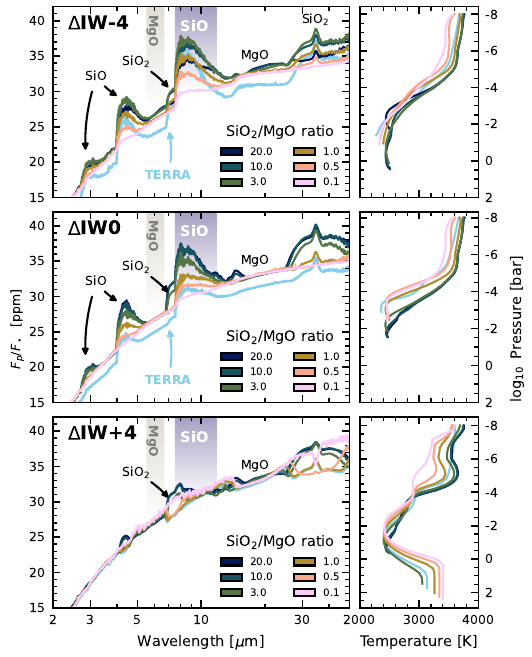}
    \caption{Planet-to-star flux ratio spectra of atmospheres above a binary MgO-SiO$_2$ melt, with varying mass ratios of MgO/SiO$_2$ (shown in inset). We tested the conditions $T_{\text{irr}}=2500$ K for an Earth-sized planet at a) $\Delta$ IW-4, b) IW and c) $\Delta$ IW+4.}
    \label{fig:mgsi_series}
\end{figure}

\begin{figure}[t]
    \centering
    \includegraphics[width=\linewidth]{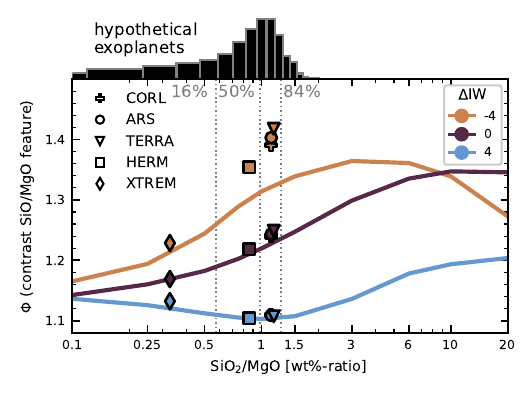}
    \caption{$\Phi$ as function of composition for a binary SiO$_2$-MgO melt (lines, colored by \fOtwo), assuming $T_{\text{irr}}=2500$ K, $1 M_\oplus$ and $1 R_\oplus$, compared to $\Phi$ for a atmosphere-free, pure black body planet (black dashed line). The $\Phi$ of the multicomponent-melts of the archetypes in Table \ref{tab:archetypes} are overlaid (markers). The distribution of SiO$_2$/MgO of hypothetical exoplanet mantles is indicated as histogram (top), with the 16, 50 and 80 percentile shown as grey dotted lines.}
    \label{fig:mgsi_feature_strength}
\end{figure}

In Section \ref{results:spectra:effect_of_comp}, it was pointed out that both increasing $a$SiO$_2$(l) and decreasing \fOtwo will result in more intense emission from SiO(g) features. To quantify the relative importance of these two variables, we ran some simple models of MgO-SiO$_2$ liquids at various compositions along the binary join, together with their respective atmospheric gas species, Mg, MgO, Si, SiO, SiO$_2$, O and O$_2$. The planet was again assumed to be Earth-sized, irradiated at 2500 K, and \fOtwo was sampled at $\Delta$IW-4, IW, and $\Delta$IW +4. The resulting MIR-spectra are displayed in Fig. \ref{fig:mgsi_series}, where they are compared to the native spectrum of TERRA (blue) at the given conditions. As expected, the strength of the SiO feature correlates with the SiO$_2$ content of the melt; the SiO(g) emission features at 9 {\textmu}m at any $T$ or \fOtwo are near-absent ($<$ 5 ppm) in SiO$_2$-poor compositions (SiO$_2$/MgO $< 1$), and become more intense for atmospheres above more silica-rich melts. MgO shows no clear features in its spectral bands, contrary to what was seen in Sec. \ref{results:spectra:effect_of_fO2} and Fig. \ref{fig:spectra_all}. Since it acts as a near-grey absorber, its "feature" in the MgO-band - as observed in the TERRA spectrum - is due to TiO(g), which acts to cool the lower atmosphere where MgO is spectrally active (see Fig. \ref{fig:mgsi_series}, right column; the $P$-$T$-profile of TERRA in blue compared to the TiO-free binary cases), leading to less thermal emission \citep{Zilinskas:2022}. 

To quantify the intensity of the SiO(g) feature, we introduce the contrast ratio $\Phi$ by comparing the intensity of its strongest feature at 9 {\textmu}m to that of an adjacent MgO band at 6 {\textmu}m (purple and grey fields in Fig. \ref{fig:mgsi_series}). We define the contrast ratio as:

\begin{equation}
    \label{eq:feature_strength}
    \Phi = \frac{\int_{\lambda_{\text{SiO},0}}^{\lambda_{\text{SiO},1}} F_p \diff \lambda}{\int_{\lambda_{\text{MgO},0}}^{\lambda_{\text{MgO},1}} F_p \diff \lambda} \cdot \frac{\int_{\lambda_{\text{MgO},0}}^{\lambda_{\text{MgO},1}} F_\star \diff \lambda}{\int_{\lambda_{\text{SiO},0}}^{\lambda_{\text{SiO},1}} F_\star \diff \lambda}
\end{equation}
where $\lambda_{\text{SiO},0}$ denotes the lower end of the SiO-band, and $\lambda_{\text{SiO},1}$ the upper (likewise for MgO). Here, we use $\lambda_{\text{MgO},0}=5$ {\textmu}m and $\lambda_{\text{MgO},1}=6.7$ {\textmu}m, $\lambda_{\text{SiO},0}=7.5$ {\textmu}m and $\lambda_{\text{SiO},1}=10$ {\textmu}m.

As shown in Fig. \ref{fig:mgsi_feature_strength}, the contrast ratio $\Phi$ at constant $T_{\text{irr}}$ is sensitive to both \fOtwo and composition. We recover the earlier discussed suppression of the SiO(g) feature in oxidising atmospheres of a fixed composition, but also find that they can still be forced into emission by increasing the SiO(g) content, which occurs when the lava has a SiO$_2$/MgO ratio significantly higher than the bulk mantle (as indicated by growing $\Phi$, see Fig. \ref{fig:mgsi_feature_strength}). Contrarily, the reducing endmember (\dIW{-4}) shows a turnover for high SiO$_2$/MgO ratios ($\geq 3$), which is the result of a growing inversion due to SiO(g), cooling the photosphere (cf. Fig. \ref{fig:mgsi_series}).

At fixed \fOtwo (i.e., along the colored lines in Fig. \ref{fig:mgsi_feature_strength}), $\Phi$ varies by a maximum amount of $\sim$0.20 over the entire range of SiO$_2$/MgO studied here (cases \dIW{-4} and \dIW{0}, while \dIW{+4} shows a significantly smaller $\Delta \Phi$), amounting to $\Delta F_p/F_\star \sim 5.4$ ppm for a $1 R_\oplus$ planet (21.53 ppm for $2 R_\oplus$). However, over the range of plausible bulk compositions - estimated by the 16th and 84th percentile in the hypothetical exoplanet SiO$_2$/MgO ratio - we observe a limited contrast ratio change of maximally $\sim 0.07$ (achieved by \dIW{-4}), amounting to $\Delta F_p/F_\star \leq 1.9$ ppm for a 1 $R_\oplus$ planet or $\leq 5.88$ ppm for 2 $R_\oplus$. This means that its more challenging to separate between the majority of expected exoplanet bulk compositions, but it is possible to identify outliers (provided \fOtwo is known), such as the SiO$_2$ depleted XTREM or (hypothetical) SiO$_2$-enriched surfaces derived from partial melting of the bulk mantle \citep{Zilinskas:2022}, e.g., SiO$_2$/MgO$\sim5.7-7$ for mid-ocean ridge basalts \citep{klein2003geochemistry} and SiO$_2$/MgO$\sim16.6$ for continental crust \citep{wedepohl1995composition}.

A change in \fOtwo at fixed SiO$_2$/MgO=0.98 (the 50\% percentile) from \dIW{-4} to \dIW{+4} induces variation of $\Delta \Phi \sim 0.21$ ($\sim 5.65$ ppm) for an Earth-sized planet or $\Delta F_p/F_\star \sim 22.6$ ppm for $R_p=2 R_\oplus$. For higher SiO$_2$/MgO ratios, we find equal or larger $\Delta \Phi$ and $\Delta F_p/F_\star$, while for lower SiO$_2$/MgO ratios, the $\Delta$'s become smaller.  Therefore, the effect of \fOtwo on $\Phi$ exceeds that of composition by a factor of at least 2 for the plausible range of mantle compositions.

Taken together, this highlights \fOtwo as a strong diagnostic tool, yet it is degenerate with composition in terms of their effect on $\Phi$, which is only exacerbated once compositions with SiO$_2$/MgO outside the main bulge of the distribution (Fig. \ref{fig:mgsi_feature_strength}, top) are considered. In order to break the degeneracy, spectral features of other species, such as SiO$_2$(g) should be investigated. Equation \ref{eq:f_of_gas_species} shows that $p$SiO$_2$ is independent of \fOtwo, but relates to the activity (and hence concentration) of SiO$_2$ in the melt. In Fig. \ref{fig:mgsi_series}, we see that its features at 7.5 and 34 {\textmu}m are indeed sensitive to the SiO$_2$/MgO ratio; unfortunately, the variation with melt composition within the 7.5 {\textmu}m feature is comparatively weak, while the (more reliable) 34 {\textmu}m feature is beyond the reach of modern telescopes, including JWST.

Above considerations were made for a binary SiO$_2$-MgO liquid, while real melts also incorporate other oxides; therefore, we compare the contrast ratios of the planetary archetypes to the simple binary liquid (markers in Fig. \ref{fig:mgsi_feature_strength}). We find great agreement for the intermediate to oxidising atmospheres, but highly reducing cases ($\Delta$IW-4) show larger contrast ratios for multicomponent silicate liquids than in the binary. This is related to the aforementioned effect of TiO(g), which weakens MgO emission, therefore enhancing $\Phi$ (see Fig. \ref{fig:mgsi_series}, and \ref{results:spectra:effect_of_fO2} and Sec. \ref{apx:TiO} for more detail). Additionally, TiO(g) seems to increase the degeneracy between melt SiO$_2$ and \fOtwo, as shown by the steep rise of $\Phi$ for the archetypes (markers in Fig. \ref{fig:mgsi_feature_strength}) compared to the pure MgO-SiO$_2$ liquid. However, a multicomponent melt releases additional redox sensitive species into the atmosphere, for example Fe(g) and TiO(g) that could help to constrain \fOtwo (see Sec. \ref{results:spectra:effect_of_fO2}), and, by lifting the degeneracy, the SiO$_2$ content of the melt as well.

\subsection{A link between melt chemistry and redox state}
\label{discussion:link_between_melts_and_redox}

At a given temperature and pressure, composition and \fOtwo are linked through the ratio of the activities of two oxidation states of the same element (e.g., $a$FeO/$a$Fe$_2$O$_3$, $a$Fe/$a$FeO or $a$Si/$a$SiO$_2$, and so on). In our model, the redox-sensitive species are Fe(l), FeO(l) and Fe$_2$O$_3$(l). 

Fe$_2$O$_3$(l), like FeO(l), is a component in the silicate melt and becomes abundant ($>$ 10 \%) only above $\Delta$ IW+6 at 2500 K, and thus has minimal influence on the atmospheric $P$-$T$-profile or spectrum (see Appendix \ref{apx:bse_fugacity_series}). However, at reducing conditions, Eq. (\ref{eq:IW_buffer}) predicts that oxidised iron, FeO(l), should no longer be stable in large quantities, instead forming metallic iron Fe(l) + 0.5O$_2$(g). We treated Fe(l) as a distinct, pure phase where $a_{Fe}=1$, which fixes the partial pressure of Fe(g) over the lava (Sec \ref{methods:vapour}, \ref{results:vapour_chemistry}). 
In reality however, the metal phase may not be pure. As oxygen fugacity decreases, iron melt may also incorporate silicon (or other elements in low quantities) and, at high temperatures, oxygen \citep[e.g.,][]{oneill1998}. The incorporation of Si into metal is predicted, based on 1 bar thermodynamic data, to occur at highly reducing conditions (below $\Delta$IW-6), but SiO$_2$ reduction is favoured with increasing $T$ \citep{oneill1998,gessmann2001,ricolleau2011}. Reduction of SiO$_2$ into Si would lower the SiO$_2$ concentration of the silicate melt; once a fraction of SiO$_2$ is reduced to its metallic form, $Si(l, metal) + O_2 \leftrightarrow SiO_2(l, silicate)$ (with activity $a$Si in the metal), it must hold that $p$Si $\propto a$Si, $p$SiO $\propto a$Si$ \cdot f\text{O}_2^{1/2}$ and $p$SiO$_2 \propto a$Si$ \cdot f\text{O}_2$. If we were to treat the formation of Si-metal in a manner analogous to Fe-metal (i.e., it forms its own, separate metallic phase), then $a$Si$=1$ and we would find constant vapour pressures of Si(g), while $p$SiO and $p$SiO$_2$ would drop for increasingly reducing conditions. This would leave the vapour increasingly dominated by Mg(g) as conditions become more reducing. In reality, Si and Fe-metal would form an alloy, and the activity of Si will be lower than 1 \citep{lacazesundmann1991} and non-trivially interlinked with $a$SiO$_2$. Simultaneously, the activity of Fe in the metal phase would likely fall as more foreign species are dissolved into the metal. This would reduce the partial pressure of iron in the vapour, leaving atmospheres sourced from highly reducing melts increasingly iron poor.

However, we acknowledge that our model cannot capture the formation of a multi-component liquid alloy as of yet. Further studies on the thermodynamics of silicate melts have to be performed, and the activities of more species should be calibrated.

Should metal formation occur in a magma ocean, as envisaged herein, then its high density relative to silicate liquid promotes its segregation and sinking to form a metallic core. Thus, the system is no longer chemically closed, and the iron-loving elements are extracted from the magma ocean. Hence, the compositions HERM and XTREM  (Table \ref{tab:archetypes}) should be descriptive of planetary melts under highly reducing conditions ($\Delta \text{IW} \leq -2$).

\subsection{Implications for geochemical characterisation of lava ocean planets}
\label{discussion:archetype_spectra}
\begin{figure}[t]
    \centering
    \includegraphics[width=\linewidth]{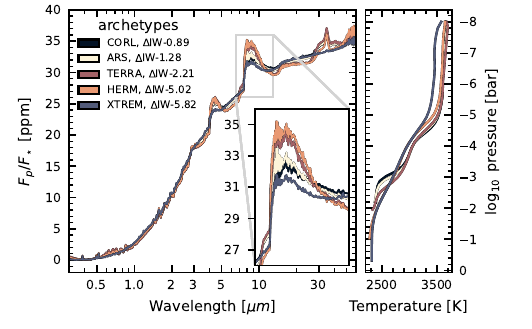}
    \caption{Secondary eclipse depth and atmospheric profiles produced by the planet/melt compositions from Table \ref{tab:archetypes}, evaluated with the respective \fOtwo (Table \ref{tab:estimated_fugacity}).}
    \label{fig:archetypes_spectra}
\end{figure}

Thus far, the mantle composition and oxygen fugacity have been treated as independent variables, though they may be linked as established above. Since the planetary archetypes in Table \ref{tab:archetypes} were constructed with a fixed total amount of oxygen, we can place a crude estimate on \fOtwo set during core-formation of these hypothetical endmembers (see Appendix \ref{apx:GMM} for more details).  The resulting fugacities are shown in Table \ref{tab:estimated_fugacity} and were used to generate atmospheric structure and spectra of said composition-\fOtwo combinations for $T_{\text{irr}}=2500$ K and 1 $R_\oplus$. Results are shown in Fig. \ref{fig:archetypes_spectra}.

\begin{table}[t]
    \centering
    \caption{Estimated oxygen fugacities at core formation of respective exoplanet compositions.}
    \label{tab:estimated_fugacity}
    \begin{tabular}{rrrrrrr}
    \toprule
    {} &   CORL &    ARS &  TERRA &   HERM &  XTREM \\
    \multicolumn{1}{l}{\textit{mantle}} & & & & & \\
    \midrule
    SiO$_2$ &  31.26 &  37.15 &  44.89 &  41.54 &  21.30 \\
    FeO     &  35.43 &  22.82 &   7.60 &   0.00 &   0.00 \\
    \multicolumn{1}{l}{\textit{core}} & & & & & \\
    \midrule
    Fe      &    &   100 &   100 &   93 &   74 \\
    Si      &    &   0 &   0 &   7 &   26 \\
    \midrule
    $\Delta$IW & $\geq{-0.89}$ & -1.30 & -2.21 & -5.02 & -5.82 \\
    \bottomrule
    \multicolumn{6}{p{0.95\linewidth}}{Notes. All quantities are in wt\%. CORL, due to being coreless, has no associated core composition.}
    \end{tabular}
\end{table}

We find the intensity of the SiO(g) feature to increase in the order XTREM-CORL-ARS-TERRA-HERM, which is mostly dictated by the descending \fOtwo. XTREM is the only exception; it expresses the weakest SiO(g) feature despite its highly reducing nature ($\Delta$IW-5.82). This is a consequence of the previously established effects of SiO$_2$/MgO-ratio and \fOtwo: the extremely low SiO$_2$ in XTREM archetype leads to a low $p$SiO. HERM also has its SiO(g) feature slightly diminished (cf. Fig. \ref{fig:spectra_all}), but due to its more reducing nature it expresses the strongest feature overall. TERRA only shows a slightly weaker feature; this echoes the degeneracy seen between low SiO$_2$ and oxidising conditions (Sec. \ref{discussion:degeneracy_between_meltcomp_and_fO2} and Fig. \ref{fig:mgsi_feature_strength}). However, Whether compositions with SiO$_2$/MgO as low as XTREM are fully molten is uncertain due to the prevalence of MgO, which has a high melting temperature of 3098 K \citep{Dubrovinsky1997}. If it is excluded, we find that, despite the degeneracy, the strength of the SiO feature relates to \fOtwo and rightfully identifies the most reducing compositions HERM and TERRA as such, while the more oxidised ARS and CORL show correspondingly weaker features.

However, the composition-fugacity combinations shown here may not be fully reflective of the actual exoplanet population. In the Solar System, Mercury has a higher SiO$_2$ abundance than predicted from solar abundances despite its low \fOtwo of $\sim$IW-5.4 \citep{Namur_2016_mercury_lava_redox}, a combination that would result in enhanced SiO(g) features. Mercury-like compositions do however still occur in the stellar compositional range with the star-to-planet composition conversion employed (see Fig. \ref{fig:archetypes}). Additionally, the fugacities in Table \ref{tab:estimated_fugacity} are expected for the core-mantle interface; the mantle - provided it contains FeO - may become more oxidised towards the surface due to evolution of the redox gradient throughout the mantle \citep{armstrong2019, Deng2020}. This effect is exacerbated for coreless planets, which should be of oxidising nature \citep{ElkinsTanton_2008}. FeO-free (reducing) mantles are less well studied, but the redox conditions might be set by species such as sulfur or carbon, as could have been the case on Mercury \citep{Namur_2016_mercury_lava_redox}. In this case, the partial pressures of O$_2$ and mineral gases can still be predicted properly by using \fOtwo as independent variable, even though the atmospheric composition and spectra would correspondingly change \citep[cf.][]{Jaeggi:2021}.

\subsection{Determination of the oxygen fugacity of exoplanet targets with MIRI}
\label{discussion:jwst}

\begin{figure*}
    \centering
    \includegraphics[width=\textwidth]{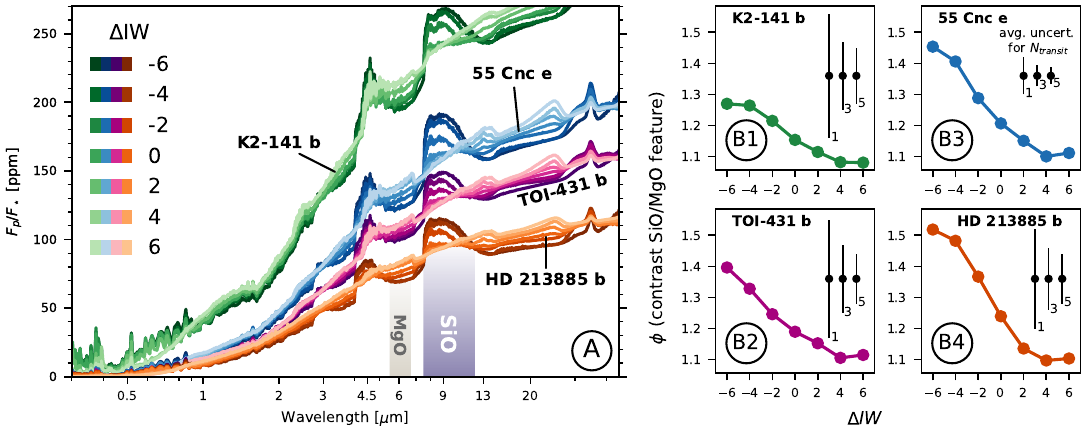}
    \caption{Emission spectra simulated for various existing lava planets, K2-141-b (green), 55-Cnc-e (blue), TOI-431-b (purple) and HD 213885-b (orange), assuming constant composition (TERRA, Table \ref{tab:archetypes}) and their corresponding stellar spectra \citep{Zilinskas:2022}. Two diagnostic features for \fOtwo determination are highlighted: MgO at 6 {\textmu}m and SiO at 9 {\textmu}m. The contrast between the fluxes of the respective bins are shown on the right, for each planet, respectively, once as the contrast expected from a perfect observation (solid coloured lines) and the respective standard deviations (1$\sigma$) for various number of observed transits with MIRI \citep[black, simulated with \texttt{pandexo}][]{Batalha_2017_pandexo}.}
    \label{fig:lava_planets_spectra}
\end{figure*}

\begin{table*}[]
    \centering
    \tiny
    \caption{Physical and orbital parameters of the four lavaplanets investigated in Fig. \ref{fig:lava_planets_spectra}.}
    \label{tab:lavaplanet_params}
    \begin{tabular}{lrrrrrrrrrrrr}
    \toprule
     & $M_p$ [$M_\oplus$] & $R_p$ [$R_\oplus$] & p [d] & T$_\text{day}$ [K] & $M_\star$ [$M_\odot$] & $R_\star$ [$R_\odot$] & T$_	\text{eff}$ [K] & Sp. T& d [pc] & magJ & magK & ESM \\
    name &  &  &  &  &  &  &  &  &  &  &  \\
    \midrule
    K2-141 b & 4.97 & 1.51 & 0.28 & 2680 & 0.71 & 0.68 & 4570 & K7V & 61.87 & 9.09 & 8.40 & 21.72 \\
    TOI-431 b & 3.07 & 1.28 & 0.49 & 2373 & 0.78 & 0.73 & 4850 & K$^a$ & 32.56 & 7.30 & 6.72 & 24.47 \\
    55 Cnc e & 7.99 & 1.88 & 0.74 & 2495 & 0.91 & 0.94 & 5172 & G8V & 12.34 & 4.59 & 4.01 & 108.28 \\
    HD 213885 b & 8.83 & 1.75 & 1.00 & 2712 & 1.07 & 1.10 & 5978 & G & 48.00 & 6.81 & 6.42 & 21.38 \\
    \bottomrule
    \end{tabular}
\end{table*}

As has been shown by \citet{Zilinskas:2022} and \citet{piette2023rocky}, the SiO$_2$(g) and SiO(g) features at 7 \& 9 {\textmu}m can be observed with JWSTs mid-infrared instrument (MIRI). In order to quantitatively examine the potential to constrain oxygen fugacity from such an observation, we construct synthetic mineral atmospheres from the evaporation of a bulk silicate Earth-like mantle composition (TERRA), together with the orbital and physical parameters of K2-141 b, 55 Cancri e, TOI-431 and HD 231885 b; all potential lava planets \citep[although recent JWST observations hint at a CO-CO$_2$ atmosphere on 55 Cnc e,][]{hu2024secondary}. Stellar spectra were taken from \citet{Zilinskas:2022}\footnote{\url{https://github.com/zmantas/LavaPlanets}}, and the physical and orbital parameters can be found in Table \ref{tab:lavaplanet_params}. We test \fOtwo between -6 and 6 log units relative to IW, assuming no heat redistribution, i.e. the dilution factor $\mathfrak{f}=\frac{2}{3}$ \citep{Hansen:2008}. As close-in planets are most likely tidally locked, this is a reasonable assumption.

The resulting spectra are shown in Fig. \ref{fig:lava_planets_spectra} A and analysed in Fig. \ref{fig:lava_planets_spectra} B in terms of the flux contrast between the MgO and SiO bands, $\Phi$ (eq. \ref{eq:feature_strength}). This provides merely a simplistic approach; actual observations should be analysed by atmospheric retrievals instead \citep[e.g.][]{piette2023rocky}, but we expect them to mainly rely on the contrast between the MgO and SiO feature to constrain the redox state.

The simulated spectra in Fig. \ref{fig:lava_planets_spectra} A show that the wavelength-dependent secondary eclipse depth varies according to the planetary and stellar characteristics, predominantly the relative size between planet and star, as well as their respective temperatures (cf. Eq. \ref{eq:flux_ratio}). In particular, the model for planet K2-141 b shows large flux ratios in the MIR (up to $\sim$250 ppm), despite its moderate size of 1.51 $R_\oplus$ as it orbits a small and cooler K-type star (roughly $4000 \leq T_\star \leq 5000$, see Table \ref{tab:lavaplanet_params}). TOI-431 b, orbiting a similar star, is the smallest and coldest planet in this sample, yet it still produces an appreciable secondary occultation depth for identical reasons as K2-141 b. 55 Cnc e produces a strong signal due to its large size, while the similarly sized HD 213885 b - the hottest planet in this sample - orbits a star slightly larger and hotter than the Sun, and thus shows a weaker signal overall.

To distinguish between various redox states, however, the relative intensity of SiO-MgO emission features (the contrast ratio, $\Phi$) is most exacting. The value of $\Phi$ shows a near-monotonic decrease as a function of \fOtwo at constant temperature and composition, for all planets (Fig. \ref{fig:lava_planets_spectra} B). Although K2-141 shows the deepest secondary occulation depth, its maximum $\Phi$ is significantly lower compared to the other planets considered (Fig. \ref{fig:lava_planets_spectra} B1); TOI-431 b also exhibits a reduced contrast ratio (Fig. \ref{fig:lava_planets_spectra} B2), yet not as extreme. This is attributable to the weaker thermal inversion expected in mineral atmospheres of planets orbiting K-stars \citep[][and c.f. Fig. \ref{fig:lava_planets_pTprofiles}, this study]{Zilinskas:2022}. These weakened inversions are the result of less short-wave radiation being available to drive heating of the upper and intermediate atmosphere, where the SiO feature emits. The planets around the "Solar-type" stars 55 Cnc and HD 213885 have stronger atmospheric thermal inversions and therefore produce a more marked contrast ratio between the SiO and the MgO features (Fig. \ref{fig:lava_planets_spectra} B3 \& B4).

To determine the extent to which such contrast ratios are observable, we construct mock observations with JWST's MIRI instrument in LRS mode using \texttt{pandexo} \citep{Batalha_2017_pandexo}, assuming the observations are stacked after 1, 3 and 5 occultations. The precision with which \fOtwo can be inverted depends not only on the flux ratio, but also the absolute flux, for which the observation of the brightest target (55-Cnc-e) results in the smallest uncertainties per occultation (Fig. \ref{fig:lava_planets_spectra} B3). In this (best-case scenario), 5 occultations would permit distinction between $\Delta \text{IW}$ cases to a precision of $\sim \pm$ 1 log unit, whereas for all other planets (B1, B2 and B4), only highly reducing ($\leq$ \dIW{-2}) or highly oxidising ($\geq$ \dIW{+4}) conditions could be ruled out.

The uncertainty on the feature strength strongly relates to the emission-spectroscopy-metric \citep[ESM, ][]{Kempton_2018}:
\begin{equation}
    \label{eq:ESM}
    \text{ESM} = 4.29 \cdot 10^6 \cdot \frac{B_{7.5}(T_\text{day})}{B_{7.5}(T_\star)} \cdot \left( \frac{R_p}{R_\star} \right)^2 \cdot 10^{-m_K/5}
\end{equation}
where $B(T)$ is the Planck function, $T_\star$ the temperature of the host star, $T_{\text{day}}$ the dayside temperature of the planet (i.e., assuming $\mathfrak{f}=\frac{2}{3}$), $R_p$ and $R_\star$ are the radii of planet and star, respectively, and $m_K$ is the K-band magnitude. It captures the quality of a target for observation, in this case with the JWST. As can be seen in Table \ref{tab:lavaplanet_params}, 55 Cnc e has by far the highest ESM, and thus requires fewer occultations to reach a given signal-to-noise ratio. All other targets have smaller ESMs, and thus larger uncertainties in Fig. \ref{fig:lava_planets_spectra}. The controlling factor is the distance of the target to the observer; the ESM scales proportional to $10^{-m_K}/5$ \citep{Kempton_2018}, introducing a variance of a factor $\sim 6$ between 55 Cnc e and the other planets. The other parameters in Eq. {\ref{eq:ESM} are not as significant as they are similar between all planets, highlighting the importance of observing nearby targets.

\subsection{Beyond JWST}
\label{discussion:beyond_JWST}

Within the atmospheres explored, we find that, in addition to the intensity of the 9 {\textmu}m SiO feature relative to the MgO background at $\sim$6 {\textmu}m, the Mg-MgO pair as well as the lines of TiO and Fe in the visible hold promise for determination of \fOtwo. These lines correspond to a high spectral emittance of the planet (c.f. Fig. \ref{fig:spectrum_fO2}) but low flux-ratio. However, the use of high-resolution cross-correlation spectroscopy (HRCCS) might prove fruitful in detecting element-oxide pairs in the atmosphere as well as species not included in this study that (potentially) have lines in the visible and are abundant in the melt: Cr-CrO-CrO$_2$, Mn-MnO, Fe-FeO, Ni-NiO,  Al-AlO or Ca-CaO, for example. Some of these species have already been observed with HRCCS to occur in Hot Jupiters \citep{hoeijmakers2019}. The relevant wavelength range of 0.3--1 micron is within the optical window of Earth's atmosphere and is therefore less contaminated by telluric lines \citep[e.g.][]{Smette_2015_molecfit}. To study the redox state of the outgassing melt, it is imperative to measure the abundances of the gases; while many studies in HRCCS aim solely at detection of species, they also allow the retrieval of atmospheric abundances \citep{Brogi:2019}.

\subsection{Assumptions and Limitations}
\label{sec:assumptions_and_limitations}

In order to complete this study, a plethora of assumptions and simplifications had to be made, listed here in the subjective order from most to least severe:

\begin{enumerate}[wide, labelwidth=!, labelindent=0pt]

    \item We treat the atmosphere as a 1D gas-column with single irradiation temperature. However, the hemispherical asymmetry implied by tidal locking could cause a continuous temperature decrease away from the substellar point, such that there is no single atmospheric temperature \citep{Zieba:2022}. Further, we cannot account for temperature variations caused by day-to-night winds \citep{castan2011atmospheres, Nguyen:2021, Nguyen:2022}. However, the impact of geometry or atmospheric circulation on emission spectra is beyond the scope of this study.
    
    \item We lack the opacity of one of the main atmospheric species, FeO(g), therefore neither the models shown here nor elsewhere \citep{Ito:2015, Zilinskas:2022, piette2023rocky} are able to account for its potential effect on atmospheric structure and spectra. Additionally, it may reveal another channel through which to derive \fOtwo by comparing the intensities of the spectral features generated by Fe(g) and FeO(g), or to link the spectra to mantle FeO content.

    \item Pressure broadening coefficients for the species considered in this study are not well understood. When available, broadening coefficients are only valid for a background atmosphere composed of hydrogen, helium or air \citep{Tennyson:2016}. However, ignoring this effect might induce larger errors. We discuss this in more detail in Section \ref{app:broadening}, where we find that the intensities of the SiO features are reduced by up to $5.36$ ppm at the SiO-9{\textmu}m feature (roughly 50 \% relative, compare with Fig. \ref{fig:spectra_all}). Further, the atomic lines of all elemental species (Fe, Mg, Si, Ti, O) are pressure-insensitive in our model. For the highly abundant Fe(g), this implies that its lines are likely saturated, and will not respond linearly to changes in atmospheric iron abundance, hindering the retrievability of the planets iron content. Systematic studies of line broadening in different background atmospheres should be undertaken.

    \item Melts at high oxygen fugacities ($\Delta \text{IW}\geq 4$) produce thick atmospheres ($> 1000$ bar) that are oxygen rich, and, in extremely hot cases (3500 K irradiation temperature) they may even reach the GPa ($10^4$ bar) range. At such pressures, our assumption of the ideal gas law is no longer be correct, requiring an equation of state to properly calculate fugacities \citep[e.g.,][]{belonoshkosaxena1991}.

    \item More accurate models for the thermodynamics of liquid solutions are required. \texttt{MAGMA} might under-predict the activities of certain melt species and hence their partial pressure above the lava. On the other hand, \texttt{VapoRock} \citep{Wolf:2022} applies a more realistic thermodynamic model, but was not calibrated on compositions foreign to terrestrial rocks.

    \item In our models (Section \ref{methods:atmospheric_speciation}), we assume a well-mixed atmosphere (in terms of elemental abundances) despite the lack of thermal convection. However, diffusion could homogenise the atmosphere. Diffusion rates in a binary gas are well approximated by the Chapman-Enskog equation \citep{chapmancowling1990}, which shows that the diffusivity of a dilute species in an atmosphere is proportional to 1/$P$ ($P$ = total pressure). Diffusion is therefore faster at lower pressures and hence more efficient at higher altitudes. The spectroscopically active regions are typically below 0.1 bar, so diffusion should mix the relevant atmospheric layers relatively well. On the other hand, the lower atmosphere is frequently cooler than the upper atmosphere (Fig. \ref{fig:p_T_profiles}), impeding mixing, either by advection or by diffusion, owing to higher $P$.

    \item We assume that the melt can be considered as a well-mixed, effectively infinite reservoir. This implies that its chemical composition and oxygen fugacity does not change during vaporisation, a condition that might be violated by finite (or unmixed) melts. This may arise once the amount of vapour can no longer be considered infinitesimally small compared to the amount of liquid. Depending on T$_{\text{irr}}$ and \fOtwo, the minimal required depth of the lava ocean can be crudely estimated as ranging from nil to $\sim$ 1000 km (see Appendix \ref{app:can_lavaocean_buffer}). The depths required will be correspondingly higher if the masses of other atmospheric gases (e.g., H, C) are present \citep{Charnoz:2023}.

    \item We assume no condensation. Contrary to more reducing atmospheres, oxidised ones above \dIW{+2} develop significant cold traps in lower layers (cf. Fig. \ref{fig:p_T_profiles}, which could induce condensation and formation of cloud layers. Cold traps may lead to cloud formation if the $P$-$T$ curve for a given condensation reaction intersects that of the atmosphere \citep{Mahapatra:2017, Herbort:2020}, and the atmospheric $P$-$T$ profile might change in response. The net effect would be to trap easily condensable species in the lower atmosphere, leaving only the most volatile (e.g., K or Na) to diffuse into the upper atmosphere. Furthermore, hot lower layers with overlying cold strata are prone to vertical convection, which we do not consider in this study. 
    
\end{enumerate}


\section{Summary and conclusions}
\label{sec:summary}

We have devised a coupled framework for silicate melt vaporisation (a modified version of \texttt{MAGMA} that allows oxygen fugacity to be set as an independent variable), gas speciation (\texttt{FastChem}) and radiative transfer (\texttt{HELIOS}) to self-consistently compute the atmospheric structure and emission spectrum of lava ocean planets (LOPs). This model was used to test how (hypothetical) exoplanet temperatures, compositions and oxygen fugacities influence atmospheric structure and resultant UV-, optical and infrared spectra. The sensitivity of spectral features in the forward models were then compared to mock observations of four LOPs, in order to identify spectral features that, through inversion, may be used to place constraints on terrestrial exoplanet geochemistry. Our key findings are the following:

\begin{itemize}

    \item The lava oceans redox state substantially shapes atmospheric chemistry. Under reducing conditions (below $\sim \Delta$IW), the atmosphere is dominated by metal-bearing gases such as SiO, Mg and Fe, while above $\Delta$IW+2,  O and O$_2$-rich atmospheres predominate. The partial pressure of Fe(g)  remains constant under strongly reducing conditions ($\leq \Delta$IW+2) due to the formation of metallic Fe. Gaseous TiO, a strong absorber, is mostly found in reducing atmospheres. Redox-neutral molecules such as MgO(g), SiO$_2$(g) and FeO(g) have constant partial pressures for all \fOtwo for a given temperature and composition, and occur in comparatively low abundance.

    \item The total atmospheric pressure scales approximately as p $\propto$~\fOtwo$^{-0.5}$ for $\Delta$IW $\leq 1$, and p $\propto$~\fOtwo for $\Delta$IW $\geq 1$. This implies that very reducing ($\Delta$IW $\leq -4$) and very oxidising ($\Delta$IW $\geq 4$) systems produce thicker atmospheres, sometimes in excess of 1 bar, while a well-defined pressure minimum occurs around IW to $\Delta$IW+2, depending on temperature. Total atmospheric pressures generated at a given $T$ and \fOtwo depend only weakly on composition, unless the planet is cold ($\leq 2500$ K) and alkali metals are present.

    \item Atmospheric $P$-$T$ profiles are nearly independent of melt composition, but are sensitive to \fOtwo. Reducing atmospheres show strong thermal inversions, resulting in hot upper atmospheres and increasing SiO emission thereby hindering cloud formation. Oxidised atmospheres have less pronounced inversions, to the point were some extreme cases become nearly isothermal, e.g. $\Delta$IW $\geq 6$ and $T$ = 3500 K). The latter can also form cold traps in their photospheres, potentially sufficient to induce cloud formation.

    \item The primary spectral features in the mid-infrared include SiO peaks at 4.5 and 9 {\textmu}m and the MgO "grey" background around 6 {\textmu}m. Both \fOtwo and the SiO$_2$ content of the melt influence their relative intensity, positively impacting SiO(g) emission. However, within plausible mantle SiO$_2$ ranges, \fOtwo-induced variation surpasses composition-induced changes. TiO(g) further decreases emission in the MgO band under reducing conditions, intensifying the contrast between SiO(g) and MgO(g) features, a trend amplified with increasing melt TiO$_2$ content.

    \item A subordinate factor influencing the intensity of SiO(g) mission is pressure-broadening. Hence, precise broadening theories, presently lacking, are essential for correlating atmospheric spectra with geochemical features of the underlying planet.

    \item The oxygen fugacity can be determined via observations made in the MIR. The diagnostic feature sensitive to \fOtwo is the ratio between the line strength of SiO at $9 \mu$m and the emission in the MgO-band $\sim 6 \mu$m, which could constrain \fOtwo $\sim\pm$ 1 log unit with JWST observations, depending on the brightness of the target and the number of occultations. Most promising for this endeavour are hot ($\geq 2500$ K) super-Earths orbiting nearby Sun-like stars (i.e., those with a high emission spectroscopy metric).

    \item Mineral atmospheres express strong lines in the visible due to evaporating neutral metal atoms, such as Fe, Mg and Si, as well as some molecules like TiO and MgO. They also carry the imprint of the melts \fOtwo and could be potentially identified by ground based telescopes through high-resolution cross-correlation spectroscopy (HRCCS).
    
\end{itemize}

\noindent \textit{Data availability.} The \texttt{MAGMA} code can be obtained from Bruce Fegley Jr. upon reasonably request. The other codes used in this study are distributed under permissive software licences and can be obtained from the respective sources (see below). The model pipeline can be found at \url{https://github.com/ExPlanetology/phaethon}. The atmospheric structure, chemistry and spectra data can be found at \url{zenodo}.\\

\noindent \textit{Acknowledgements.} We are grateful to Laura Schaefer for a rigorous and constructive review that resulted in substantial improvements to the work. We appreciate the editorial handling and input of the editor, Emmanuel Lellouch. We thank Bruce Fegley Jr. for providing the \texttt{MAGMA} code, Lukas Carmichael for providing the exoplanet composition data and Matteo Brogi for our discussions on the possibility of HRCCS for small exoplanets. This work was supported by the Swiss National Science Foundation (SNSF) through an Eccellenza Professorship (203668) and the Swiss State Secretariat for Education, Research and Innovation (SERI) under contract No. MB22.00033, a SERI-funded ERC Starting grant "2ATMO" to P.A.S. Parts of this work have been carried out within the framework of the National Centre of Competence in Research (NCCR) PlanetS supported by the SNSF under grant 51NF40\_205606. This publication makes use of The Data \& Analysis Center for Exoplanets (DACE), which is a facility based at the University of Geneva (CH) dedicated to extrasolar planets data visualisation, exchange and analysis. DACE is a platform of the Swiss NCCR PlanetS, federating  Swiss expertise in Exoplanet research. The DACE platform is available at \url{https://dace.unige.ch}. This work made use of the following codes: \texttt{HELIOS} \citep{Malik:2017, Malik:2019}, \texttt{FastChem} \citep{Stock:2018}, \texttt{Pandexo} \citep{Batalha_2017_pandexo}, \texttt{MAGMA} \citep{Fegley:1987, Schaefer:2004}, \texttt{VapoRock} \citep{Wolf:2022}, \texttt{numpy} \citep{numpy_harris2020array}, \texttt{scipy} \citep{scipy_2020SciPy-NMeth}, \texttt{pandas} \citep{pandas_mckinney-proc-scipy-2010}, \texttt{matplotlib} \citep{matplotlib_Hunter:2007}, \texttt{seaborn} \citep{seaborn_Waskom2021} and \texttt{scikit-learn} \citep{scikit-learn}.

\bibliographystyle{aa.bst} 
\bibliography{bibliography}


\begin{appendix}

\section{Elaborations on outgassing}
\label{apx:bse_fugacity_series}

Here, we treat the outgassing from a Bulk Silicate Earth (BSE) composition \citep{mcdonough1995}, spanning the range of $\Delta$IW-6 to $\Delta$IW+8 to highlight \textit{a.)} the effect of alkalis on the atmospheric pressure and composition and \textit{b.)} the evolution of the concentration and activity of FeO and Fe$_2$O$_3$ as a function of redox state.

\begin{figure}[!t]
    \centering
    \includegraphics[width=\linewidth]{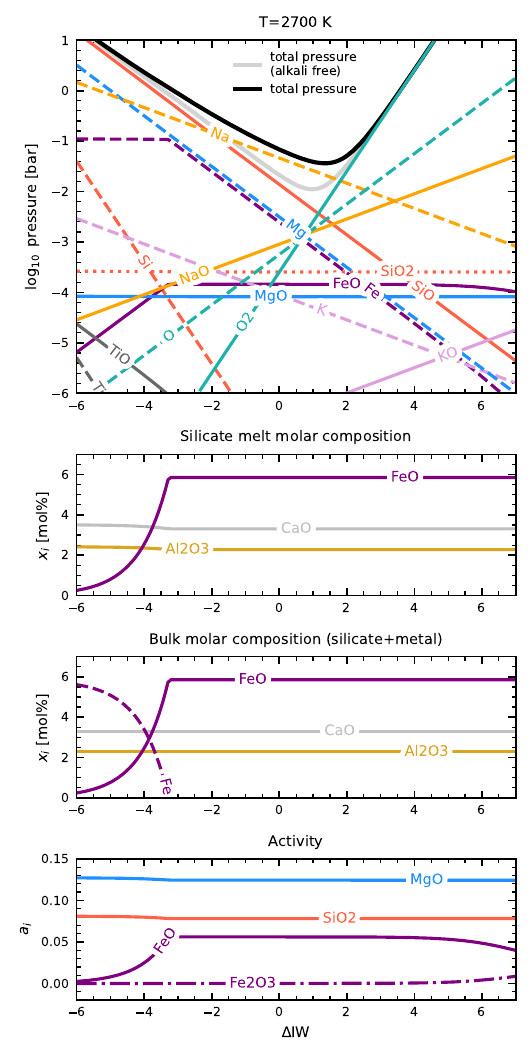}
    \caption{Vapour pressures and thermodynamics of a silicate melt of BSE composition. \textbf{Upper:} Similar to Fig. \ref{fig:melt_vapor_evolution}, but for a melt of BSE composition (including Na$_2$O). The dominance of Na(g) at intermediate redox states (\dIW{-3} to \dIW{+2}) is evident, which is consistent with the findings of other studies \citep{Schaefer:2004, Miguel:2011, Wolf:2022, vanBuchem:2023}. \textbf{Upper center:} Concentration of oxides in the silicate melt. FeO concentration deplets under reducing conditions, as Fe is formed. This is evident in \textbf{Lower center:} here, the bulk system composition is shown, highlighting the conversion of FeO to Fe. \textbf{Lower:} Activity of Fe and Fe$_2$O$_3$ in the melt, compared to the far more dominant SiO$_2$ and MgO.}
    \label{fig:BSE_fugacity_series}
\end{figure}

\subsection{Alkali metals}
\label{apx:bse_fugacity_series:alkali}

As can be seen in the upper panel of Fig. \ref{fig:BSE_fugacity_series}, Na(g) is prevalent under reducing and highly reducing conditions (\dIW{$\leq 0$}), but reacts less sensitively to \fOtwo than SiO(g) (cf. Eq. \ref{eq:f_of_gas_species} and Table \ref{tab:coeffs}). Hence, it becomes the dominant species in atmospheres of reducing to intermediate redox states ($\Delta$IW-2 to +2), significantly increasing the total atmospheric pressure compared to the alkali free case. This effect diminishes as temperature increases, as SiO(g) is more temperature sensitive \citep[][ and Sec. \ref{results:vapour_chemistry:temp} this study]{Wolf:2022}. The increase in total pressure can be seen in the atmospheric profiles in Fig. \ref{fig:alkali_test}. Under oxidising and highly oxidising conditions (\dIW{$\geq +2$}), Na(g) and NaO(g) are the more relevant metal gas compared to species of Si, Mg and Fe. As was seen in Fig. \ref{fig:alkali_test}, Na(g) maintains some spectral features in the visible+NIR even under oxidising conditions, which is contrary to e.g. Fe(g) and Mg(g) (cf. Fig. \ref{fig:spectrum_fO2}). Potassium K(g) and KO(g) do not play significant roles in terms of abundances or spectra for BSE-like abundances (cf. Fig. \ref{fig:alkali_test}).

\subsection{Iron speciation and melt thermodynamics}
\label{apx:bse_fugacity_series:thermochemistry}

As described in Sec. \ref{methods:vapour}, the speciation of iron changes with \fOtwo; both within the silicate melt phase as Fe$^{3+}$ and Fe$^{2+}$, and as Fe$^{0}$, which forms a new phase because it is insoluble in silicate melts. Consider a chemical system with \fOtwo as an independent variable (Fig. \ref{fig:BSE_fugacity_series}). In such a system, the total moles of Fe, Mg, Al, Ca, Na, K, Ti and Si are constant as a function of \fOtwo, while the number of moles of O is permitted to vary in accordance with \fOtwo to satisfy Eqs. \ref{eq:IW_buffer} and \ref{eq:WH_buffer}. When \fOtwo is sufficiently low that the system is saturated in Fe$^{0}$ (i.e., $x$Fe > 0, as given by Eq. \ref{eq:IW_buffer}), the concentration (mole fraction) of FeO in the melt must decrease to satisfy the constraint that $\sum$Fe = constant (Eq. \ref{eq:mass_balance}, see Fig. \ref{fig:BSE_fugacity_series}, lower panel). In so doing, the mole fractions of other oxides in the silicate melt increase by a factor $(1-x\text{Fe}_{\text{new}}) / (1-x\text{Fe}_{\text{old}})$. This is observable in the small increases in $x$CaO and $x$Al$_2$O$_3$ in Fig. \ref{fig:BSE_fugacity_series} for a BSE-like composition. This effect would be exacerbated for compositions initially richer in FeO, like ARS or CORL. We emphasise that, although their abundances in the silicate melt phase increase, the moles of all the elements in the bulk system (except O) are conserved (Fig. \ref{fig:BSE_fugacity_series}, lower center); FeO is simply replaced by Fe as \fOtwo decreases, thereby maintaining the same number of moles of non-oxygen elements in the system.

In terms of activities, $a$FeO is low under reducing conditions due to the aforementioned formation of Fe-metal (see Sec. \ref{methods:vapour}), but increases as conditions become more oxidising. The activity of FeO stabilises once Fe-metal destabilizes. This transition point is temperature and composition dependent, and for the BSE composition at 2500 K located at $\sim \Delta$IW-3. This value is too low; for these conditions, the transition should be closer to $\Delta$IW-2 \citep{Frost:2008a_redox_coreformation}, likely a consequence of the crude thermodynamic model applied by \texttt{MAGMA}. $a$FeO shows a second "roll-over" at very high \fOtwo, at which more and more FeO is converted into Fe$_2$O$_3$ (indicated by the increase of the Fe$_2$O$_3$ activity, and decrease of $a_{FeO}$). This suppresses the outgassing rate of iron, as shown in Fig. \ref{fig:BSE_fugacity_series}, where FeO(g) shows a roll-over at very oxidising states ($\geq \Delta$IW+6). This would yield very oxidising atmopsheres increasingly iron poor. However, it remains to be investigated if planets with such oxidising mantles can form, and if they could maintain such massive O$_2$-atmospheres in excess of 1 GPa.

\FloatBarrier

\section{\texttt{FastChem} reactions}
The reactions allowed in \texttt{FastChem} are listed in Table \ref{tab:fastchem_reactions}; this includes all species that can form in the mineral atmospheres of this study. The "basic" reactions cover the chemistry available in all atmospheres, while the "extended" selection is only used in specific use cases, i.e. Sec. \ref{discussion:alkali}. Species that also have an associated opacity and outgass significantly are highlighted in bold. Al, AlO, Ca and CaO for example would also be available opacity sources, but outgass in such low mixing ratios that we do not consider them. FeO however, despite its prevalence in mineral atmospheres, lacks line lists and therefore cannot be included.


\begin{table}[!h]
\centering
\caption{Gas species considered in our atmospheric chemistry model (see Sec. \ref{methods:atmospheric_speciation}).}
\label{tab:fastchem_reactions}

\begin{tabular}{|c|l|l|}
\hline
& \textbf{element} & \textbf{species} \\
\hline
\multirow{7}{*}{basic}                        & Si & \textbf{Si}, \textbf{SiO}, \textbf{SiO$_2$}, Si$_2$, Si$_3$, Si$^+$, Si$^-$                              \\
                                              & Mg & \textbf{Mg}, \textbf{MgO}, Mg$_2$, Mg$^-$                                              \\
                                              & Fe & \textbf{Fe}, FeO, Fe$^+$, Fe$^-$                                              \\
\multirow{2}{*}{}                             & Al & Al, AlO, AlO$_2$, Al$_2$, Al$_2$O, Al$_2$O$_2$, \\
                                              &     & AlO$^+$, Al$^-$, AlO$^+$, AlO$_2^-$, Al$_2$O$^+$ \\
                                              & Ca & Ca, CaO, Ca$_2$, Ca$^+$, Ca$^-$                                         \\
                                              & Ti & \textbf{Ti}, \textbf{TiO}, TiO$_2$, Ti$^+$, Ti$^-$                                        \\
                                              & O  & \textbf{O}, \textbf{O$_2$}, O$_3$, O$^+$, O$^-$ , O$_2^+$, O$_2^-$                                   \\
                                              & e$^-$ & e$^-$ \\
\hline
\multicolumn{1}{|l|}{\multirow{2}{*}{ext.}} & Na & \textbf{Na}, NaO, Na$_2$, NaO$^-$, Na$^+$, Na$^-$                                   \\
\multicolumn{1}{|l|}{}                          & K  & \textbf{K}, K$^+$, K$^-$, KO, K$_2$, KO$^-$                                         \\
\hline
\end{tabular}
\end{table}

\FloatBarrier

\section{Determination of hypothetical exoplanet compositions}
\label{apx:GMM}

To derive the hypothetical exoplanet compositions we utilise (Table \ref{tab:archetypes}), we use a Gaussian mixture model (GMM), which can identify distinctive archetypes within the population, as depicted in Figure \ref{fig:archetypes}. GMMs characterize a given population of $d$-dimensional datapoints by approximating it with a distribution composed of $k$ $d$-dimensional Gaussians, with $d$ being the number of parameters. For the problem at hand, it is best to transform the composition of mantle (wt\% of SiO$_2$, MgO, CaO, Al$_2$O$_3$ and FeO) and core (wt\% of Fe and Si) into a logratio w.r.t to MgO (mantle) or Fe (core) in order to remove the restriction of having to sum to 100\%, while the core-mass-fraction CMF was also $\log$-transformed in order to avoid the boundaries at 0 and 1. A challenge emerges from the oversimplified condensation model, which produces a trifurcation in the resultant planetary classifications: while 'Earth-like' planets (e.g., ARS, TERRA) exhibiting a well-defined parameter set, coreless planets are characterized by an absence of a core and its composition (e.g., CORL in our dataset), whereas planets formed under reducing conditions (HERM \& XTREM) lack mantle FeO. This implies that $d$ varies between the subclasses, and we therefore have to treat each one individually before collecting the final sample of exoplanet archetypes.

While the choice of $k$ is arbitrary, one can ascertain an "optimal" number by varying it and evaluating the Bayesian Information Criterion (BIC) \citep{Schwarz:1978:BIC}:

\begin{equation}
\label{eq:BIC}
\text{BIC} = -2 \log \hat{L(k)} + \log N \cdot d
\end{equation}

Here, $N$ represents the number of samples, and $\hat{L}$ denotes the likelihood as determined by the respective algorithm in use \citep[in this case, the \texttt{GaussianMixtureModel} from \texttt{scikit-learn},][]{scikit-learn}. Typically, one would anticipate the BIC to be minimized for the number of Gaussian components that best fits the data without overfitting. However, the distribution of hypothetical exoplanets is not distinctly Gaussian, rendering the BIC less decisive in identifying an optimal $k$. Instead, we primarily examine the gradient of the BIC and identify points where it flattens, signifying a potential optimal number of Gaussian components.

We use $k=2$ for coreless and reduced, FeO-free planets and $k=6$ for regular terrestrial planets with an iron core and mantle FeO. The determination of these node counts was informed by the earlier analysis of the BIC, although we use less components to fit the reduced sub-population than predicted by consideration of the BIC (5 components). In total, we obtain 10 Gaussian modes, each characterized by a multidimensional mean representing a descriptive exoplanet composition.

From the resulting set of 10 potential candidates, we filter out compositions that either have a low weight (contributing minimally to the overall population) or are too similar to another composition in the set (e.g., marginal difference in their mantle chemistry). This process yields a final sample of five planets that exhibit maximum diversity. These compositions, detailed in Table \ref{tab:archetypes}, are named according to their characteristics: CORL represents the population of coreless planets; TERRA serves as an analog for Earth; ARS, richer in mantle iron but with a smaller core than Earth, can be likened to a "Super-Exo-Mars"; HERM, fully depleted in FeO yet with a relatively high SiO$_2$/MgO ratio, bears semblance to Mercury albeit with a smaller core; and XTREM, unlike any observed in the Solar System, is extremely reduced, leading to Si entering the core rather than the mantle, thereby enriching the mantle with MgO.

Additionally, with the core composition (Table \ref{tab:archetypes}), we are able to estimate the oxygen fugacity at core-mantle equilibrium (Table \ref{tab:estimated_fugacity}). To do so requires assumptions to be made as to the equilibrium conditions at the core-mantle interface. Here we assume core formation takes place at a fictive p-T; 2000 K and 1 bar, at which the thermodynamic properties of the pure endmembers \citep[Fe, Si;][]{chase1998nist} and of the mixtures \citep[Fe-Si alloys;][]{lacazesundmann1991} are well known, facilitating calculation of \fOtwo. Depending on the Fe-Si-O element balance, \fOtwo is determined accordingly:

\begin{enumerate}
    \item The ratio of FeO(silicate) to Fe(metal) sets the fugacity at the core-mantle boundary:
    \begin{equation}
        \label{eq:Delta_IW}
        \Delta \text{IW} = 2 \log \left( \frac{\gamma_{FeO} \cdot x_{FeO}}{\gamma_{Fe} \cdot x_{Fe}} \right)
    \end{equation}
    following the law of mass action. We set $\gamma_{FeO}=1.3$ \citep{oneill2002}, $\gamma_{Fe}=0.9$ which accounts for the dilution of pure Fe with Ni, which is reduced at higher \fOtwo than is Fe.

    \item The subsample of the planets that have insufficient oxygen to oxidise all of the Si present in the bulk planet need to be treated separately, since some Si partitions into the metallic core. Hence, the oxygen fugacity is set to the Si-SiO$_2$ (dubbed here "Silicon-Quartz" buffer, abbrev. SiQ):
    \begin{equation}
        \label{eq:Delta_SiQ}
        \Delta \text{SiQ} = \log \left( \frac{\gamma_{SiO_2} \cdot x_{SiO_2}}{\gamma_{Si} \cdot x_{Si}} \right)
    \end{equation}
    We set $\gamma_{SiO_2}=0.8$ \citep{Wolf:2022}, $\gamma_{Si}=0.25$ for XTREM and $\gamma_{Si}=0.01$ for HERM \citep{lacazesundmann1991}.
    The \fOtwo given by eq. \ref{eq:Delta_SiQ} can be expressed relative to the IW-buffer by adding a correction factor ($\Delta IW$ at 2000 K, 1 bar):
    \begin{equation}
        \Delta \text{IW} = \Delta \text{SiQ} - 7.327
    \end{equation}

    \item Some planets in our sample might be coreless. This occurs if there is sufficient oxygen to oxidise all Fe, as is the case for CORL, then the FeO/FeO$_{1.5}$ ratio dictates \fOtwo \citep[e.g.,][]{oneill2018}. The simplified condensation model does not account for the formation of Fe$_2$O$_3$. As such, we use eq. (\ref{eq:Delta_IW}) to provide a minimum estimate for the \fOtwo at which the silicate liquid would be in equilibrium with pure Fe. For this planet archetype, it yields $\Delta$IW $\geq$ -0.89.
    
\end{enumerate}

\section{Effect of TiO on spectrum}
\label{apx:TiO}

\begin{figure}[t]
    \centering
    \includegraphics[width=\linewidth]{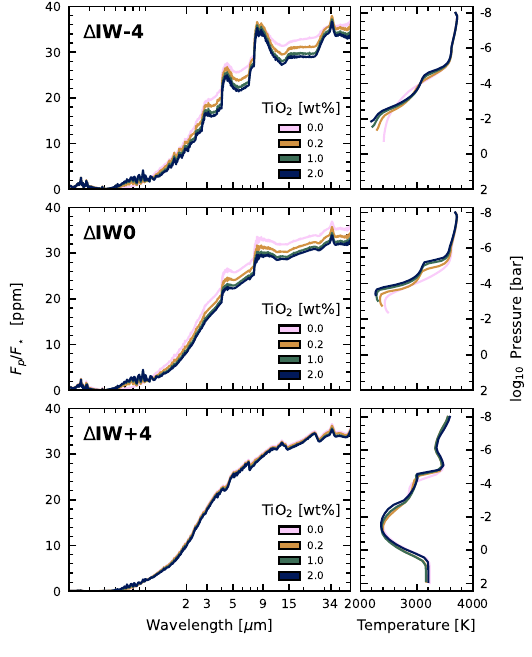}
    \caption{Spectra of planets with enhanced TiO$_2$ melt content. The underlying melt composition is assumed to be TERRA, and the TiO$_2$ melt abundance is enhanced according to the main text. The planet under study is Earth-sized and irradiated at 2500 K.}
    \label{fig:tio2_series}
\end{figure}

TiO(g) is a strong shortwave absorber and thus might modify atmospheric spectra significantly, hence posing as a potential probe for melt composition. To test its effect, we used the composition TERRA and variend the TiO$_2$ abundance from 0.1 to 5 wt\%. Results are displayed in Fig. \ref{fig:tio2_series}. As can be seen, increasing levels of TiO$_2$ produce stronger emission in the visible and near-infrared, but reduce emission in the mid-infrared globally. TiO predominantly absorbs/emits in the range 0.45 - 1 micron, coinciding with maxima in stellar emissions for G-type stars, thus it absorbs stellar radiation before it reaches lower layers. 
As a consequence, the contrast between the MgO and SiO bands is enhanced as the lower atmosphere - where MgO resides - is cooled. Reducing systems ($\Delta$IW-4) show lower emission in MgO(g) and SiO(g) at 4.5 {\textmu}m, but SiO(g) at 9 {\textmu}m is not as strongly affected, possibly a combination of the 9 {\textmu}m feature being more opaque (cf. Fig. \ref{fig:molecular_opacities}) and SiO(g) being more abundant.  This enhances the contrast $\Phi$ (Eq. \ref{eq:feature_strength}) with increasing melt TiO$_2$ content for this redox regime. Intermediate conditions ($\Delta$IW 0) do not adjust the shape of their spectral features to the TiO$_2$ abundance, but lower overall emissivity. Oxidising conditions (here $\Delta$IW+4) suppress outgassing of TiO(g), therefore they are hardly affected.

However, TiO$_2$ is generally not expected to occur in large abundances in planetary mantles due to its overall low availability in the protoplanetary nebula \citep{lodders1998planetary}; it can be, however, accumulated at the surface via fractional melting or crystallisation due to its nature as an incompatible element. The lunar mare basalts for example can contain several wt\% of TiO$_2$ \citep[up to $\sim 13$ wt\%,][]{Giguere:2000_lunar_basalt}. On Earth, crustal material (formed through fractional melting of the mantle) is enriched in SiO$_2$ and TiO$_2$ while depleted in MgO, which would generate stronger SiO features (Sec. \ref{results:spectra}). As already noted by \citet{Zilinskas:2022}, such a combination could be used to distinguish evaporating bulk mantle from crustal material.

\FloatBarrier

\section{Stellar spectra influence on pressure-temperature profile}

The major cause for the prominent inversions found earlier is the absorption of shortwave radiation (VIS and NIR), and reemission in MIR. As G-type stars - which we assumed as the fiducial case in our model - have their emission maximum in this wavelength regime, they readily cause thermal inversions \citep{Zilinskas:2022}. However, some of the planets tested in Sec. \ref{discussion:jwst}, namely K2-141 b and TOI-431 b, orbit K-type stars. The spectral maxima of these stars lies deeper in the infrared, thus limiting the UVIS input to their planets atmosphere. As a consequence, the thermal inversion of mineral atmospheres around cooler stars are less pronounced (see Fig. \ref{fig:lava_planets_pTprofiles}), and the SiO(g) feature is weakened \citep{Zilinskas:2022}. This implies that retrieval of \fOtwo, as outlined in Sec. \ref{discussion:jwst}, becomes more challenging for planets around cooler stars.

\begin{figure}[t]
    \centering
    \includegraphics[width=\linewidth]{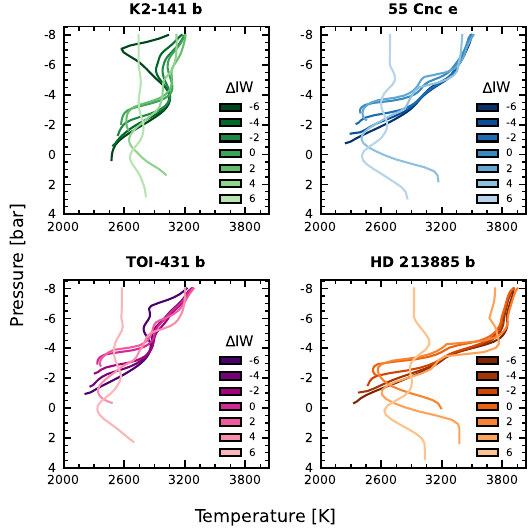}
    \caption{Pressure-temperature profiles of atmospheres in Fig. \ref{fig:lop_representative_atmospheres}. }
    \label{fig:lava_planets_pTprofiles}
\end{figure}

\FloatBarrier

\section{Effect of pressure broadening}
\label{app:broadening}

Even though the broadening coefficients used in this study are estimated values, neglecting broadening altogether would result in even less accurate results. To quantify the extent of the broadening effect as a function of \fOtwo, we repeat an analysis for a 1 $R_\oplus$, 1 $M_\oplus$ planet with $T_{irr}=2500$ K and the TERRA composition for the fugacity range in $\Delta$IW from -6 to +6, shown in Fig. \ref{fig:without_pressurebroadening}. The opacities were constructed as outlined in Section \ref{methods:opacities:molecules}, but restricted at a single pressure value of $p=10^{-8}$ bar. Temperature (Doppler) broadening is allowed.
The unbroadenend spectra (thick lines) show significantly smaller SiO(g) features, but remain similar to the broadened lines (dashed), indicated by the difference between the respective cases. Pressure-temperature profiles change only noticeably for the highly oxidised cases $\Delta$IW +4 and +6. 

\begin{figure}[!h]
    \centering
    \includegraphics[width=\linewidth]{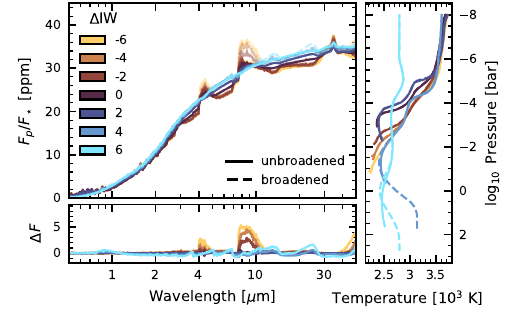}
    \caption{Spectra of a bulk silicate Earth-like (TERRA) planet of 1 $R_\oplus$ (as per Fig. \ref{fig:spectra_all}) and 2500 K when no pressure broadening is considered (bold lines) vs. the broadened spectra (dashed) seen in the main paper (see Fig. \ref{fig:spectra_all}). The unbroadenend lines are sampled at a constant pressure of $10^{-8}$ bar. The SiO features are significantly less pronounced when pressure broadening is ignored.}
    \label{fig:without_pressurebroadening}
\end{figure}

\FloatBarrier
\section{Can the lava ocean buffer the atmosphere?}
\label{app:can_lavaocean_buffer}

We assumed that the melt can be considered as an infinite reservoir compared to the vapour. To investigate this assumption, we compare the column density for an (assumed isothermal) atmosphere:
\begin{equation}
    N_{O_2, atmo} = \frac{p_{O_2}}{m_{O_2} \cdot g}
\end{equation}
where $p$ is the pressure (Pa), $m$ the mass of the average atmospheric particle (kg) and g the gravitational acceleration (m/$s^2$), with the column density of a melt-layer:
\begin{equation}
    N_{O_2, melt} = \frac{\rho h}{m_{melt}} \cdot x_{O_2}
\end{equation}
where $\rho$ is the density of the melt (kg/$m^3$), $h$ is the depth of the melt column (m), and $m$ is its average molecular weight ($kg$), and $x_{O_2}$ being the molar fraction of O$_2$ in the melt. Approximate values for each of these parameters are: $m_{atmo}=32$, $m_{melt}$ based on melt composition but typically $\sim 51$ g/mol, $\rho \sim 2600$ kg/m$^3$ and $g \sim 10$ m/$s^2$, $p_{O_2}$ is based on the evaporation model at given temperature and \fOtwo conditions. $x_{O_2}=0.25$, based on its rough stoichiometry in the melt (which is by mole number half composed of O). The melt composition is assumed to be BSE (bulk silicate Earth, similar to the TERRA composition in our study); $m_{melt}$ is therefore $\sim 51.9 \cdot u$, where $u$ is the atomic mass unit.

By requiring that $N_{O_2, melt} \geq N_{O_2, atmo}$, we can estimate the depth of the melt column that can buffer the atmosphere, labeled here as the minimal necessary ocean depth (MNOD). MNOD is obtained by finding the value $h$ for which $N_{O_2, melt} = N_{O_2, atmo}$; therefore, the actual ocean depth has to exceed MNOD by at least an order of magnitude. Results are displayed in \ref{fig:MNOD}.

\begin{figure}[t]
    \centering
    \includegraphics[width=\linewidth]{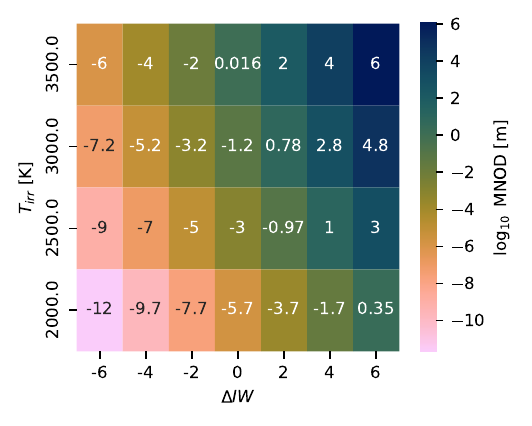}
    \caption{Minimal necessary ocean depth (MNOD) to buffer the atmosphere, in meters.}
    \label{fig:MNOD}
\end{figure}

While this estimate is incredibly simplistic and qualitative at its best, it still indicates that \textit{a.)} cooler melts should be more capable of buffering the atmosphere, given the low column density of the vapour and \textit{b.)} oxidising lava oceans must be of greater depth than reducing ones in order to buffer the vapour \fOtwo, based on the large column density of the massive O+O$_2$ atmospheres.

\end{appendix}

\end{document}